\documentclass[fleqn,usenatbib]{mnras}

\usepackage{newtxtext,newtxmath}

\usepackage[T1]{fontenc}

\DeclareRobustCommand{\VAN}[3]{#2}
\let\VANthebibliography\thebibliography
\def\thebibliography{\DeclareRobustCommand{\VAN}[3]{##3}\VANthebibliography}

\usepackage{graphicx}	
\usepackage{amsmath}	
\usepackage{caption}
\usepackage{subcaption}
\usepackage{float}
\usepackage[table]{xcolor}
\newcommand{\rulesep}{\unskip\ \vrule\ }

\title[A Machine Learning search for CVs in Gaia]{Machine Learning based search for Cataclysmic Variables within Gaia Science Alerts}

\author[D. Mistry et al.]{
D. Mistry,$^{1}$\thanks{}
C. M. Copperwheat,$^{1}$
M. J. Darnley$^{1}$
and I. Olier$^{2}$
\\
$^{1}$Astrophysics Research Institute, Liverpool John Moores University, IC2, Liverpool Science Park, 146 Brownlow Hill, Liverpool, L3 5RF, UK\\
$^{2}$School of Computer Science and Mathematics, Liverpool John Moores University, James Parsons Building, 3 Byrom Street , Liverpool, L3 3AF, UK
}

\date{Accepted XXX. Received YYY; in original form ZZZ}

\pubyear{2022}

\begin{document}
\label{firstpage}
\pagerange{\pageref{firstpage}--\pageref{lastpage}}
\maketitle

\begin{abstract}
Wide-field time domain facilities detect transient events in large numbers through difference imaging. For example, Zwicky Transient Facility, produces alerts for hundreds of thousands of transient events per night, a rate set to be dwarfed by the upcoming Vera Rubin Observatory. The automation provided by Machine Learning (ML) is therefore necessary to classify these events and select the most interesting sources for follow-up observations. Cataclysmic Variables (CVs) are a transient class that are numerous, bright and nearby, providing excellent laboratories for the study of accretion and binary evolution. Here we focus on our use of ML to identify CVs from photometric data of transient sources published by the Gaia Science Alerts program (GSA) - a large, easily accessible resource, not fully explored with ML. Use of light curve feature extraction techniques and source metadata from the Gaia survey resulted in a Random Forest model capable of distinguishing CVs from supernovae, Active Galactic Nuclei and Young Stellar Objects with a 92\% precision score and an 85\% hit rate. Of 13,280 sources within GSA without an assigned transient classification our model predicts the CV class for $\sim$2800. Spectroscopic observations are underway to classify a statistically significant sample of these targets to validate the performance of the model. This work puts us on a path towards the classification of rare CV subtypes from future wide field surveys such as the Legacy Survey of Space and Time.
\end{abstract}




\section{Introduction}

Over the last few decades, the increasing depth and breadth of imaging produced by wide field survey facilities has brought forth a revolution in time domain astronomy. Their ability to rapidly and repeatedly image huge areas of the sky combined with the technique of difference imaging, in which the new sky image is subtracted from a reference image, has greatly increased the discovery potential of time varying sources \citep{RN134}. Surveys include the Catalina Real-time Transient Survey (CRTS; \citealt{RN256}), Palomar Transient Factory (PTF; \citealt{RN405}), Panoramic Survey Telescope and Rapid Response System (Pan-STARRS; \citealt{RN406}), Zwicky Transient Facility (ZTF; \citealt{RN68}), and All-Sky Automated Survey for SuperNovae (ASAS-SN; \citealt{RN52}). In addition, the space based Gaia mission \citep{RN260}, primarily purposed for astrometry, has been recognised as a powerful tool for time domain astronomy.

Difference imaging can reveal genuine astrophysical sources that have changed in brightness or position as well as artifacts posing as such. Artifacts can make up a significant proportion of events found from difference imaging, these consist of poorly subtracted galaxies, cosmic rays, PSF halos, defective pixels, and CCD edge effects \citep{RN7}. The vetting of such artifacts is no longer solely performed by human inspection, but is now reliant in large part on automated pipelines (e.g., \citealt{RN412,RN254}). The classification of astrophysical sources has historically been performed by human inspection (e.g., \citealt{RN57,RN58,RN59}), however, technological advancements in the design of telescopes, detectors, and computing over recent decades have led to a rapid rise in transient events to classify, thus fueling the desire for ever more efficient classification methods.

Currently ZTF produces up to 10\textsuperscript{6} alerts (individual photometric data points of time variable sources) per night, and in a few years’ time the Vera Rubin Observatory Legacy Survey of Space and Time (LSST; \citealt{RN60}) will be generating up to 10\textsuperscript{7} alerts per night \citep{RN292}. Only a small fraction of associated transient sources can benefit from follow-up observations due to the limited availability of dedicated facilities and the associated telescope time constraints. As a consequence, one must be selective about the sources that are followed up, these will be those which provide the greatest potential in furthering our understanding of the transient classes to which they belong. Finding these sources requires some level of initial source classification to distill the incoming alert stream to a list of such targets. An additional requirement is the need for prompt classification as early time follow-up can be highly informative, for example, spectroscopy obtained within a few days of a supernova explosion can inform different progenitor models \citep{RN466}. To address these challenges, real time transient event processing must be accommodated into survey architecture. The level of automation needed can be achieved by integration of Machine Learning (ML). Implementation of ML requires little or no human intervention beyond the training of the associated algorithms with labelled data. The resultant models are capable of rapidly classifying unseen examples, requiring only a few seconds at most to complete each task.

Survey facilities such as ZTF have adopted ML to perform tasks such as separating out real transient events from artifacts (bogus or false positives; \citealt{RN412}); and to separate spatially extended targets (such as galaxies) from stars \citep{RN261}. This use of ML helps to pare down the large influx of alerts generated by difference imaging. Alerts brokers have been used by ZTF to ingest and characterise the nature of their alerts, serving them to the astronomical community. One such broker is ALeRCE (Automatic Learning for the Rapid Classification of Events; \citealt{RN421}). The AleRCE pipeline makes use of science, reference and difference images for rapid classification of events \citep{RN419}, and multiband light curves for longer term characterisation \citep{RN420}. The pipeline uses these inputs to group events into several transient classes, such as cataclysmic variables (CV), supernovae (SN), Active Galactic Nuclei (AGN), variable stars, and young stellar objects (YSOs). Aside from use within brokers, ML has been used for classification of SN sub-classes \citep{RN281,RN362} as well as Galaxy morphology classification \citep{RN275}.

An area of time domain where ML classification will be of key importance is that of CVs (extensively covered in \citealt{RN9,RN445}), systems which are used to develop our understanding of binary evolution (e.g., \citealt{RN398,RN459,RN465}). These are semi-detached binary systems composed of a white dwarf (WD) and a late-type main sequence companion. The companion (or donor) is Roche lobe filling, leading to a transfer of mass to the WD via the inner Lagrangian point. In the majority of cases this results in the formation of an accretion disc. However, where the WD is strongly magnetic the accretion mechanism changes. In polars (or AM Her stars; \citealt{RN307}) the strong magnetic field ($\sim10\leq B\leq80 MG$) causes the accretion flow from the donor to be funneled by field lines onto one or both of the WD' magnetic poles. In intermediate polars \citep{RN313} ($\sim1\leq B\leq10 MG$) a partial accretion disk may form where only the inner regions of the disk are magnetically disrupted. CVs give rise to a plethora of observable phenomena. Examples include the highly energetic eruptions of classical novae (CNe), caused by the violent expulsion of the accreted shell of matter from the surface of the white dwarf driven by runaway thermonuclear reactions \citep{RN72}; the less energetic but much more frequent outbursts of dwarf novae (DNe) modelled by thermal/viscous instabilities in the accretion disk \citep{RN395}; and the recently identified micronovae, also believed to be thermonuclear runaway events, though localised to magnetically confined regions on the WD surface \citep{RN486, RN485}. Large area time-domain surveys are more likely to catch these events in action by monitoring larger numbers of sources.

These systems aid our understanding of the currently uncertain progenitor scenarios of type Ia supernovae \citep{RN44}, where both single and double degenerate pathways exist. Novae such as M31N 2008-12a \citep{RN62}, are the most promising single degenerate pathway, these comprise a near Chandrasekhar mass WD accreting at high rates. A double degenerate pathway may be provided by the ultra-short period (5-65 minute) helium rich CVs that make up the AM CVn subclass \citep{RN70}. CVs with known orbital periods are also important for such binary evolution studies, these enable the masses and radii of the individual stars to be determined. The orbital period can be deduced in the SU UMa dwarf nova subtypes from an eruption induced periodic ‘superhump’ variation in the light curve  \citep{RN46}; from radial velocity variations in spectral lines \citep{RN385}; or in eclipsing systems via the periodic occultation of the WD and accretion disc by the donor \citep{RN49}. The exploration of wide field survey photometry with ML techniques provides an avenue for the discovery of many more of these transients. This is an active area of research, where examples include work by \citet{RN65} to distinguish between CVs, supernovae, and several other classes from a dataset constructed from CRTS light curves; and the classification of alerts from the ZTF stream within the AleRCE alerts broker pipeline \citep{RN420}.

In this work we describe our exploration of data generated by the Gaia spacecraft \citep{RN360} to identify new members of the CV population. Gaia is now recognised as a powerful tool for transient detection, with Gaia Science Alerts (GSA; \citealt{RN358}) providing alerts of newly discovered transient sources at a current rate of $\sim$12 per day by repeatedly scanning the whole sky. The cadence of associated light curves is dictated by the `Gaia scanning law' \citep{RN360} - typically, a pair of observations separated by 106.5 minutes is separated by another pair two to four weeks later. The photometry is precise to 1\% at G=13, and 3\% at G=19. This resource therefore provides a stable platform from which to evaluate ML based classification. In Section \ref{sec:Dataset}, we describe the classified transients of GSA; the methods used to extract relevant descriptive characteristics from their light curves; and the additional metadata gathered from the survey for each source. In Section \ref{sec:Method} we describe how the resultant dataset was used to train several ML algorithms to perform a set of classification tasks, along with a description of how the resultant models can be evaluated. In Section \ref{sec:Results} we detail the performance of each algorithm. Finally, we discuss the outcomes of our exploration of GSA along with a description of a pilot study involving spectroscopic classification to validate predictions made by our best performing model (Section \ref{sec:Discussion}).

\section{Dataset} \label{sec:Dataset}
\subsection{Gaia alerts and EDR3}\label{sec:Alerts}

As of June 2021, close to 18,000 transient sources had been listed within the Gaia transient alerts stream \footnote{\url{http://gsaweb.ast.cam.ac.uk/alerts/alertsindex}}; just over 4700 of which had been assigned class labels. The classifications are based upon human inspection of Gaia data in combination with the results of positional cross-matching with the Simbad \citep{RN403}, NED and VSX databases\footnote{The NASA/IPAC Extragalactic Database (NED) is funded by the National Aeronautics and Space Administration and operated by the California Institute of Technology. VSX is the International Variable Star Index database, operated at AAVSO, Cambridge, Massachusetts, USA.}, and YSO catalogues (see section 2.7.7 of \citealt{RN358}) to identify already-confirmed transient or variable objects. This information is aided by the hourly parsing of 27 major transient survey websites for reported discoveries that also contain classification information, these include Transient Name Server (TNS)\footnote{https://www.wis-tns.org/}, CRTS, ASAS-SN and Astronomer's Telegrams\footnote{https://www.astronomerstelegram.org}. Further details regarding the alerts filtering and classification process are contained in \citet{RN358}.

The process of training and validating machine learning models requires accurate class labels. Whilst we believe the aforementioned process of class assignment can reliably provide this accuracy, an inspection of class labels for a sample of these source was performed for a level of verification. Of the 2713 supernovae, 2530 are spectroscopically confirmed according to TNS, Astronomer's Telegrams contains details of spectroscopic classification for the remainder. Of the 613 Gaia labelled CVs, 471 are associated with known/confirmed CVs according to the comments associated with the Gaia classifications. Comparison with VSX confirms this along with either a confirmation of CV status or candidate status for the remainder through references to relevant research papers and Astronomer's Telegrams. Gaia's comments associated with sources labelled as AGN and YSO show 929 of the 940 transients labelled as AGN, and 184 of the 190 transients labelled as YSOs are associated with known/confirmed AGN and YSOs respectively. This was verified for a sample of these sources by examining records within TNS and associated links (e.g., Simbad). The remaining candidate AGN and YSOs were not further considered for this work.

Our dataset is composed of features extracted from light curves of these classified targets within Gaia's alert stream along with their associated class labels. Supplementary data for these targets may be available within the database of Gaia Early Data Release 3 (EDR3; \citealt{RN449,RN450}) in the form of astrometric and further photometric data such as parallax, proper motion, and photometric colour provided by the low resolution photometry (R=100) of blue and red photometers on board Gaia. A coordinate cross match with EDR3 provides this metadata for $\sim$45\% of sources within our dataset. This metadata has also been incorporated as a set of supplementary features.

Of the 4697 classified targets incorporated into our dataset, SNe account for 58\% of classified targets, AGN make up 21\%, CVs and YSOs constitute 13\% and 3\% respectively, while microlensing, tidal disruption events, and various other classes account for the remainder. 

The majority of GSA classifications come from dedicated spectroscopic follow-up programs such as PESSTO \citep{RN472} and Spectral Energy Distribution Machine (SEDM; \citealt{RN473}) that are heavily biased towards supernova classification. The class fractions of classified targets are generally dictated by what has been chosen to be classified, with unusual or ambiguous examples often overlooked, and therefore it must be noted that these fractions may not be representative of the entire sample of GSA targets.

\subsection{Light curve feature extraction} \label{sec:LC_feat_ext}

From source light curves we extracted quantitative characteristics (or features) that describe their variability. These comprised of simple statistical and periodicity based features in Table \ref{tab:Feature_w/o_feets} along with features obtainable from the feATURE eXTRACTOR FOR tIME sERIES (feets) package \citep{RN451} based on the light curve data available (magnitude and time, no error measurements), a selection of which are shown in Table \ref{tab:feets}. They consist of statistical, periodicity, and percentile based features.

\begin{table}
 \caption{Features extracted from light curves (without {\fontfamily{qcr}\selectfont feets} package)}
 \label{tab:example}
 \begin{tabular}{lp{0.65\columnwidth}}
  \hline
  Feature & Description\\
  \hline
  \textit{mean\_mag} & Mean of magnitudes\\
  \textit{median\_mag} & Median of magnitudes\\
  \textit{std\_mag} & Standard deviation of magnitudes\\
  \textit{mad\_mag} & Median absolute deviation of magnitudes\\
  \textit{min\_mag} & Minimum magnitude (maximum brightness)\\
  \textit{max\_mag} & Maximum magnitude (minimum brightness)\\
  \textit{n\_obs} & Number of observations\\
  \textit{diff\_min\_mean} & Difference between \textit{min\_mag} and \textit{mean\_mag}\\
  \textit{diff\_min\_median} & Difference between \textit{min\_mag} and \textit{median\_mag}\\
  \textit{detected\_time\_diff} & Time span of observations\\
  \textit{n\_peaks\_rm\_x\_y} & Number of observations within a rolling window of \textit{y} observations that are brighter than \textit{x} magnitudes of the median magnitude of that window (\textit{x = 1, 2, 3, 4, or 5}, \textit{y = 7}).\\
  \textit{kurtosis} & Kurtosis of the magnitudes\\
  \textit{skew} & Skewness of the magnitudes\\
  \textit{pwr\_max} & Largest power value in the Lomb Scargle Periodogram\\
  \textit{freq\_pwr\_max} & Frequency corresponding to  \textit{pwr\_max}\\
  \textit{FalseAlarm\_prob} & Estimate of the false alarm probability given the height of the largest peak in the periodogram (see \url{https://docs.astropy.org/en/stable/api/astropy.timeseries.LombScargle.html#astropy.timeseries.LombScargle.false_alarm_probability})\\
  \hline
 \end{tabular}
 \label{tab:Feature_w/o_feets}
\end{table}

\begin{table}
    \caption{A small selection of features available from the {\fontfamily{qcr}\selectfont feets} package. The full list is available at (\url{https://feets.readthedocs.io/en/latest/tutorial.html}) along with detailed explanations. Of the full list, only those requiring a magnitude and time, or just magnitude data, were implemented here.}
    \begin{tabular}{lp{0.50\columnwidth}}
        \hline
        Feature & Description\\
        \hline
        \textit{Amplitude} & Half of the difference between the median of the maximum 5\% and the median of the minimum 5\% magnitudes\\
        \textit{AndersonDarling} & The Anderson-Darling test is a statistical test of whether a given sample of data is drawn from a given probability distribution (normal distribution)\\
        \textit{Autocor\_length} & Cross-correlation of a signal with itself\\
        \textit{Eta\_e} ($\eta^e$) & Variability index $\eta$ is the ratio of the mean of the square of successive differences to the variance of data points.\\
        \textit{FluxPercentileRatioMid\textbf{X}} & Ratio of centred flux percentile ranges. If $F_{5,95}$ is the difference between the 95th and 5th percentile of ordered magnitudes, then \textit{FluxPercentileRatioMid\textbf{X}} = ${F_{40,60}}/{F_{5,95}}$, ${F_{32.5,67.5}}/{F_{5,95}}$, ${F_{25,75}}/{F_{5,95}}$, ${F_{17.5,82.5}}/{F_{5,95}}$, and ${F_{10,90}}/{F_{5,95}}$, for \textbf{X} = 20, 35, 50, 65, and 80 respectively.\\
        \textit{Freq\textbf{i}\_harmonics\_amplitude\_j} & Amplitude of the jth harmonic of the \textbf{i}th frequency component of the Lomb Scargle Periodogram\\
        \textit{Gskew} & Median-of-magnitudes based measure of the skew\\
        \textit{LinearTrend} & Slope of a linear fit to the light-curve\\
        \textit{MaxSlope} & Maximum absolute magnitude slope between two consecutive observations\\
        \textit{Meanvariance} & Ratio of the standard deviation to the mean magnitude\\
        \textit{PairSlopeTrend} & Considering the last 30 (time-sorted) measurements of source magnitude, the fraction of increasing first differences minus the fraction of decreasing first differences\\
        \textit{PeriodLS} & Period corresponding to frequency of maximum power in the Lomb Scargle Periodogram\\
        \textit{PercentAmplitude} & Largest percentage difference between either the max or min magnitude and the median\\
        \textit{Psi\_eta} & $\eta^e$ index calculated from the phase-folded light curve\\
        \textit{SmallKurtosis} & Small sample kurtosis of the magnitudes\\
        \hline
    \end{tabular}
    \label{tab:feets}
\end{table}

\subsection{Supplementary features}
\label{sec:supplementary}



Supplementary data (or metadata) from Gaia EDR3 relating to position, photometry and astrometry are incorporated as dataset features. Positional features consist of: right ascension, declination, Galactic (and ecliptic) longitude and latitude, along with associated errors. Photometric features encompass the mean flux from the red and blue photometers (BP and RP) as well as that from G band photometry; the associated mean magnitudes; colours (BP-RP, BP-G, G-RP) and associated errors. Proper motion and parallax (along with their errors) are included as astrometric features. A full list is displayed in Table \ref{tab:EDR3}, while further details are available within the Gaia EDR3 documentation\footnote{\url{https://gea.esac.esa.int/archive/documentation/GEDR3/Gaia_archive/chap_datamodel/sec_dm_main_tables/ssec_dm_gaia_source.html}}

\begin{table}
 \caption{Supplementary data from Gaia EDR3 incorporated as dataset features (see subsection \ref{sec:supplementary})}.
 \begin{tabular}{lp{0.50\columnwidth}}
  \hline
  Feature & Description\\
  \hline
  \textit{ra, dec, ra\_error, dec\_error} & Right ascension, declination, and associated standard errors\\
  \textit{l, b} & Galactic longitude and Galactic latitiude\\
  \textit{ecl\_lon, 'ecl\_lat} & Ecliptic longitude and Ecliptic latitude\\
  \textit{bp\_rp, bp\_g, g\_rp} & BP-RP, BP-G, and G-RP colours\\
  \textit{phot\_\textbf{X}\_mean\_flux} & Mean flux in the G, integrated BP, or integrated RP bands - corresponding to \textit{\textbf{X} = g, bp, or rp} respectively\\
  \textit{phot\_\textbf{X}\_mean\_flux\_error} & Error on the mean flux in the \textbf{X} band\\
  \textit{phot\_\textbf{X}\_mean\_flux\_over\_error} & Mean flux in the \textbf{X} band divided by its error\\
  \textit{phot\_\textbf{X}\_mean\_mag} & Mean magnitude in the G, integrated BP, or integrated RP bands - corresponding to \textit{\textbf{X} = g, bp, or rp} respectively\\
  \textit{pseudocolour, pseudocolour\_error} & The astrometrically estimated effective wavenumber of the photon flux distribution in the astrometric G band, measured in $\mu^{-1} m$, and standard error of pseudocolour\\
  \textit{parallax, parallax\_error} & Gaia parallax in milliarcseconds (mas) and standard error\\
  \textit{parallax\_over\_error} & Parallax divided by its standard error\\
  \textit{pm, pmra, pmdec} & Total proper motion, and proper motion in the right ascension and declination directions (mas/year)\\
  \textit{pmra\_error, pmdec\_error} & Standard error of the proper motion in right ascension and declination directions (mas/year)\\
  \textit{ruwe} & renormalised unit weight error: expected to be around 1.0 for sources where the single-star model provides a good fit to the astrometric observations. A value significantly greater than 1.0 (say, >1.4) could indicate that the source is non-single or otherwise problematic for the astrometric solution\\
  \hline
 \end{tabular}
 \label{tab:EDR3}
\end{table}

\section{Method} \label{sec:Method}
\subsection{Machine Learning algorithms}

The dataset described above can be used to evaluate the ability of ML algorithms to identify CVs within GSA. The algorithms whose performance we evaluated are SciKit-Learn's \citep{RN458} Python implementation of Random Forest (RF) \citep{RN279}, AdaBoost (ADB) \citep{RN435}, K-Nearest neighbours (KNN) \citep{RN447}, and Support Vector Machines (SVM) \citep{RN436}. Also used are the Extreme Gradient Boosting (XGBoost) algorithm \citep{RN295} and Keras \citep{chollet2015keras} 
implementation of an Artificial Neural Network (ANN) in the form of a Multi-Layer Perceptron - a fully connected multi-layer ANN \citep{RN483}.

RF, ADB and XGBoost are implementations of an ensemble of Decision Trees \citep{RN278}. Based on the features provided, Decision Trees perform successive binary splits of the training dataset in a way that resultant groups are as different from one another as possible, and closer to a homogeneity of class. The resultant model uses this tree structure to classify unseen examples. RF classifies based on a voting system using the predictions of a random collection of Decision Trees, the class with the most votes is our model's prediction. Each tree is trained on a modified version of the original training set (Bootstrap Aggregation) and a random subset of features to introduce uncorrelated trees. ADB \citep{RN435} combines Decision Trees sequentially, weights are assigned to examples such that incorrect predictions of the tree are given higher weights than those correctly predicted. This iterative process of modifying weights before training means that difficult-to-predict examples are given more influence. XGBoost is another example of sequentially combining Decision Trees. Where ADB uses weights to improve performance, XGBoost aims to reduce some error function that describes the classification performance of successive trees \citep{RN295}.

SVMs \citep{RN436} find the ideal hyperplane that best distinguishes between two classes in feature space. For non-linear class separation, SVM uses the \textit{kernel trick}, transforming lower dimension input feature space into a higher dimensional space allowing for linear separation. KNN \citep{RN447} stores the position vectors of training set class examples in feature space. Class predictions on new examples are made by assigning the mode of the classes of the k nearest neighbours from the training set to the new example. Artificial Neural Networks (ANN) \citep{RN448} consist of layers of interconnected nodes (or neurons) - an input layer, consisting of feature values; an output layer, which delivers the predictions (e.g., class probabilities); and, in between, one or more hidden layers of neurons, which sequentially transform the feature values into the predictions by applying typically non-linear functions to linear combinations of prior inputs. The algorithm learns through a process of loss minimisation whereby the model parameters are adjusted to reach convergence to loss minimum (known as back propagation).

\subsection{Fine Tuning}

Controlling how the algorithms learn from the dataset to generate predictive models is done by tuning their hyperparameters. They are algorithm settings that are set prior to the learning process. The aim with hyperparameter tuning is to improve model performance whilst the risk of overfitting (i.e. learning the noise in the data) is reduced. The hyperparameters explored for each of the algorithms tested are as given in table \ref{tab:Hyperparameters}:

\begin{table*}
    \caption{The hyperparameters explored for each ML algorithm.}
    \centering
    \begin{tabular}{p{0.3\textwidth}p{0.65\textwidth}}
        \hline
        \textbf{RF Hyperparameters} & \textbf{Description}\\
        \hline
        \textit{n\_estimators} & Number of Decision Trees\\
        \textit{max\_features} & maximum number of features provided to each tree\\
        \textit{max\_depth} & maximum number of binary split levels in each tree\\
         & \\
        \hline
        \textbf{ADB Hyperparameters} & \\
        \hline
        \textit{n\_estimators} & Same as for RF\\
        \textit{learning\_rate} & Weight assigned to each classifier at each boosting iteration. This determines the impact of each tree on the final outcome.\\
        \textit{max\_depth} & Same as for RF\\
         & \\
        \hline
        \textbf{XGBoost Hyperparameters} & \\
        \hline
        \textit{n\_estimators} & Same as for RF\\
        \textit{min\_child\_weight} & Minimum sum of weights of all observations in a child node\\
        \textit{gamma} & Nodes are split only when there is a reduction in the error defined by a loss function. Gamma specifies the minimum loss reduction required to make a split\\
        \textit{subsample} & Fraction of examples to be randomly sampled for each tree\\
        \textit{colsample\_bytree} & Similar to \textit{max\_features} in Random Forest\\
        \textit{max\_depth} & Same as for RF\\
         & \\
        \hline
        \textbf{SVM Hyperparameters} & \\
        \hline
        \textit{Kernel} & see text: `Radial Basis Function (RBF)'\\
        \textit{Kernel Coefficient ($\gamma$)} & Defines how far the influence of a single training example reaches, where the values can be seen as the inverse of the radius of influence.\\
        \textit{Error Penalty (C)} & Controls the cost of miss-classification on the training data. Small C = soft margin, large C = hard margin.\\
         & \\
        \hline
        \textbf{KNN Hyperparameters} & \\
        \hline
        \textit{n\_neighbors} & Number of nearest neighbours to use\\
         & \\
        \hline
        \textbf{MLP Hyperparameters} & \\
        \hline
        \textit{learning\_rate} & Controls how much to change the model in response to the error each time the model weights are updated.\\
        \textit{Number of Hidden Layers} & Number of hidden layers\\
        \textit{Number of neurons} & Number of units (neurons) within a given hidden layer.\\
        \textit{Activation function} & Converts the output of a neuron into a form that serves as input for the next. Used to introduce non-linearity to a network.\\
        \hline
    \end{tabular}
    \label{tab:Hyperparameters}
\end{table*}

\subsection{Classification tasks} \label{sec:Classification_tasks}

The classes assigned by the Gaia team are not mutually exclusive. For example, quasi-stellar objects (QSOs) are extremely luminous AGN. From the Gaia assigned classes, many variations of class grouping could be put forward for ML classification algorithms to distinguish. Two such groupings are defined by the following classification tasks, listed as transient class followed by the number of dataset samples in brackets:

\begin{enumerate}
    \item \textbf{Binary classification} - CV (613) or not CV (4084).
    \item \textbf{4 class classification} - this comprises of the most populous transient types in the dataset: AGN (which includes QSOs and BL Lac) as a single class (929), CVs (613), all different supernova types (SNe; 2713) and Young Stellar Objects (YSOs; 184).
\end{enumerate}

The tasks are assigned to the ML algorithms and their performance is evaluated. The classification tasks were firstly performed with both the light curve extracted features and supplementary features. However, between 58\% and 90\% of data is missing for supplementary features. This was either due to unsuccessful cross-matching of targets with EDR3 - cross-matching was unsuccessful for 90\% of supernovae, 23\% of CVs, and < 1\% of AGN and YSOs respectively - or certain metadata not being available where cross-matching was successful. For example, parallax measurements may not be available if the target is too faint or distant for an accurate measurement. Therefore we felt it necessary to also perform classification tasks with light curve extracted features alone. These implementations can then be compared with other works where classification has been performed using light curve derived features alone.

\subsection{Data pre-processing}

Prior to ingestion into ML algorithms, the associated datasets require some level of preparation. Examples within each task specific dataset contain missing data for several features. The strategy employed here is to replace missing values with the mean value of the feature column (mean imputation \citealt{RN467}). Feature scaling is employed for all except the ensemble learning algorithms (i.e., RF, ADB, and XGBoost) so that features with a larger range of values do not impart more influence on the model during training and for faster convergence to error minimum for Gradient Descent algorithms. We standardised the data to achieve zero mean and unit variance (or equivalently, standard deviation; \citealt{RN468}).

\subsection{Train-test split}

Training and evaluation of a ML model requires a separate training and test set. The algorithms are trained on the training set to generate a model to be evaluated on the test set. The task specific datasets are split 50/50 into a training set and test set in a stratified manner - the same proportion of each class is represented in each of the training and testing sets. The split is performed before the pre-processing (imputation and feature scaling) stages to avoid information from the test set being present within the training set (data leakage) and yielding extremely biased results on model performance.

\subsection{Optimal Hyperparameter Search}

Model performance depends significantly on the selection of hyperparameters. Testing all combinations manually for the optimal combination is unfeasible. Therefore the GridSearchCV and RandomisedSearchCV functions from SciKit-learn's \citep{RN458} model\_selection Python package were used to loop through combinations of predefined hyperparameters to identify an optimal set of hyperparameters for a given model. The cross-validation method is used to evaluate the performance of each combination. Cross-validation involves randomly dividing a set of observations into k groups, or folds, of approximately equal size. For each unique fold, the algorithm is trained and a model built on k-1 folds, this model is then evaluated on the remaining fold (validation set) with a chosen performance metric. This is repeated until each of the k-folds has served as a validation set. The average of the k recorded accuracies (or other chosen metric) is the cross-validation score serving as the model performance metric. For this work a 10 fold cross-validation was implemented. The training set was entered into these functions and the best combination of hyperparameters for a given algorithm was found. The model with this combination was then evaluated on the test set. 

\subsection{Classifier Performance}

The number of true positives (TP), true negatives (TN), false positives (FP) and false negatives (FN) serve as the basis for performance metrics used to assess model performance. These quantities firstly require the definition of the positive class, this is the class of interest (CV in this case); and the negative class, non-CVs in the binary classification task, and all other classes in the multi-class case.

\subsubsection{Confusion Matrix}
It is common to present the redTP, TN, FP, and FN quantities in an N $\times$ N table known as a \textit{confusion matrix} (where N represents the number of classes). This construct allows us to easily identify the number of TPs, TNs, FPs and FNs. These values may then be used to evaluate the performance of our binary and multi-class ML models using the class specific precision, recall, and F1-score; and the overall model accuracy and balanced accuracy.

\subsubsection{Precision, Recall, and F1-score}
The Precision is defined as the fraction of examples our model predicted as belonging to the positive class that do actually belong to this class: $TP/(TP+FP)$. In other words it tells us how much we can trust our models predictions of the positive class. The Recall is the fraction of examples of the positive class that our model correctly predicted as belonging to this class: $TP/(TP+FN)$. This metric assesses the model's ability to identify all members of the positive class. The F1-score is the harmonic mean of precision and recall for our positive class, and is useful in finding the best trade-off between these quantities. The highest possible value of the F1 score is 1 (100\%), indicating perfect precision and recall, the lowest possible value (0) relates to a score of 0 for either precision or recall.

\subsubsection{Accuracy and Balanced Accuracy}
The accuracy is the fraction of all examples whose class was correctly predicted by the model. For binary classification this is $(TP+TN)/(TP+TN+FP+FN)$, for a multi-class situation we sum the number of true positives for each class and divide by the total number of examples. The accuracy returns an overall measure of the model's predictive capability. Should we only be concerned with assigning the most number of examples to their correct class, accuracy is a good metric. However, under this metric, strong classification errors for classes with few examples to their name will be hidden. Therefore, should we be concerned with finding a model which has a strong classification performance across all classes, we may use `balanced accuracy' which can account for this class imbalance. This is the arithmetic mean of the recalls for each class.

\subsubsection{Area under the Curve of the Receiver Operating Characteristic (ROC)}
The Receiver Operating Characteristic (ROC) curve allows us to see the trade off between sample purity and completeness plotted as the True Positive Rate (TPR; also called recall) as a function of the False Positive Rate (FPR; the fraction of examples incorrectly predicted as belonging to the positive class: $FP/(TN+FP)$) for all values of a threshold probability above which a positive classification is made. More specifically, for each example, algorithms return a probability of belonging to a certain class. The probability is thresholded such that examples with probabilities equal or greater than the threshold are mapped to the positive class and remaining examples are mapped to the negative class. Therefore the ROC curve is an evaluation of TPR and FPR as we continuously vary this probability threshold. This can be used to determine the appropriate threshold for a given study such that we can adjust for our desired level of purity and completeness depending on our science case. The goal of classification is to maximise the TPR while minimising the FPR. For the binary classification tasks, the Area Under the Curve (AUC) of the ROC can be used to assess model performance. A value of 1 for the AUC indicates a perfect model, capable of correctly assigning the correct class prediction to all examples, when AUC = 0.5, the model is no better than a random guess, while AUC = 0 corresponds to incorrectly predicting classes of all examples.

\section{Results} \label{sec:Results}

Tables \ref{tab:Results_Binary} and \ref{tab:Results_4class} show the model evaluation scores for the binary and 4 class classification tasks. The scores for models trained with both light curve extracted and supplementary features (full feature models) are shown without brackets, while the scores for models trained with light curve extracted features alone (light curve only models) are within brackets. The scores shown are the accuracy, balanced accuracy, and with respect to the CV subclass, the precision, recall, and F1-score. The choice of best performing model is based on the F1-score for the CV class. This metric was chosen as it considers both the need to minimise FPs, which is important for the efficient use of telescope time for target follow-up, and a requirement to minimise FNs.

\begin{table}
    \centering
    \caption{Binary task classification scores for ML models as measured on the test set. Scores without brackets relate to models using both light curve and supplementary features, while those in brackets are for models that used only light curve extracted features. Random Forest was implemented with 100, 250, 750, and 1000 trees denoted by RF then the number of trees; other abbreviation are ADA - AdaBoost, MLP - Multi Layer Perceptron, KNN - K Nearest Neighbours and SVM - Support Vector Machine}
    \label{tab:Results_Binary}
    \begin{tabular}{lccccc}
        \hline
        Model & Accuracy & Balanced & CV & CV & CV\\
        & & Accuracy & Precision & Recall & F1-score\\
        \hline
        RF100 & 0.955 & 0.870 & 0.88 & 0.76 & 0.81\\
            & (0.938) & (0.824) & (0.82) & (0.67) & (0.74)\\
        RF250 & 0.955 & 0.870 & 0.89 & 0.75 & 0.81\\
            & (0.939) & (0.829) & (0.82) & (0.68) & (0.74)\\
        RF500 & 0.955 & 0.868 & 0.89 & 0.85 & 0.81\\
            & (0.937) & (0.827) & (0.81) & (0.68) & (0.74)\\
        RF750 & 0.955 & 0.867 & 0.89 & 0.75 & 0.81\\
            & (0.937) & (0.826) & (0.81) & (0.67) & (0.74)\\
        RF1000 & 0.956 & 0.870 & 0.90 & 0.75 & 0.82\\
            & (0.938) & (0.826) & (0.82) & (0.67) & (0.74)\\
        ADA & 0.959 & 0.874 & 0.91 & 0.76 & 0.83\\
            & (0.932) & (0.840) & (0.75) & (0.72) & (0.73)\\
        XGBoost & 0.962 & 0.883 & 0.92 & 0.78 & 0.84\\
            & (0.943) & (0.823) & (0.86) & (0.67) & (0.76)\\
        MLP & 0.932 & 0.824 & 0.78 & 0.68 & 0.72\\
            & (0.932) & (0.822) & (0.78) & (0.67) & (0.72)\\
        KNN & 0.909 & 0.812 & 0.65 & 0.68 & 0.66\\
            & (0.900) & (0.812) & (0.60) & (0.69) & (0.64)\\
        SVM & 0.817 & 0.787 & 0.39 & 0.75 & 0.52\\
            & (0.871) & (0.802) & (0.51) & (0.71) & (0.59)\\
        \hline
    \end{tabular}
\end{table}

\begin{table}
    \centering
    \caption{4 class classification scores. Score with and without brackets, and abbreviations are as described in table \ref{tab:Results_Binary}.}
    \label{tab:Results_4class}
    \begin{tabular}{lccccc}
        \hline
        Model & Accuracy & Balanced & CV & CV & CV\\
        & & Accuracy & Precision & Recall & F1-score\\
        \hline
        RF100 & 0.964 & 0.941 & 0.92 & 0.85 & 0.88\\
            & (0.922) & (0.835) & (0.80) & (0.79) & (0.80)\\
        RF250 & 0.964 & 0.936 & 0.92 & 0.85 & 0.88\\
            & (0.924) & (0.835) & (0.81) & (0.79) & (0.80)\\
        RF500 & 0.965 & 0.941 & 0.92 & 0.85 & 0.88\\
            & (0.923) & (0.830) & (0.80) & (0.80) & (0.80)\\
        RF750 & 0.965 & 0.942 & 0.92 & 0.86 & 0.89\\
            & (0.923) & (0.833) & (0.80) & (0.80) & (0.80)\\
        RF1000 & 0.965 & 0.942 & 0.92 & 0.86 & 0.89\\
            & (0.923) & (0.833) & (0.80) & (0.80) & (0.80)\\
        ADA & 0.959 & 0.925 & 0.90 & 0.82 & 0.86\\
            & (0.897) & (0.798) & (0.75) & (0.75) & (0.75)\\
        XGBoost & 0.962 & 0.928 & 0.91 & 0.84 & 0.87\\
            & (0.922) & (0.820) & (0.83) & (0.77) & (0.80)\\
        MLP & 0.926 & 0.873 & 0.80 & 0.76 & 0.78\\
            & (0.910) & (0.793) & (0.73) & (0.71) & (0.76)\\
        KNN & 0.895 & 0.795 & 0.90 & 0.48 & 0.63\\
            & (0.898) & (0.730) & (0.86) & (0.67) & (0.75)\\
        SVM & 0.891 & 0.798 & 0.62 & 0.70 & 0.66\\
            & (0.874) & (0.758) & (0.76) & (0.61) & (0.68)\\
        \hline
    \end{tabular}
\end{table}

\subsection{Binary classification}
\subsubsection{Full feature model}

The best performing binary task full feature model was XGBoost, trained with 150 Decision Trees at a learning rate of 0.1 and maximum tree depth of 6. The model outperformed each of the others with an F1-score of 84\%, the AdaBoost and Random Forest implementations follow closely behind (81-83\%). There is though little difference between the top two models; use of the McNemar's test to compare the XGBoost and AdaBoost models show both classifiers make errors in much the same proportion, (for $\alpha=0.05$; $p=0.175$). The confusion matrix (left panel of figure \ref{fig:Binary_CMs}) indicates 69 of the 307 CVs in the test set were miss-classified by the XGBoost model, while of the 258 examples predicted as CVs only 20 were not. The corresponding ROC curve is plotted in the left panel of figure \ref{fig:Binary_XGB_AUC}, with AUC score of 0.975. The importance of each feature for a given model can be given by the feature importance scores. The 20 features with the largest effect on the model's predictive accuracy are plotted in the left panel of figure \ref{fig:Binary_XGB_FI}. The number of observations greater than 2 magnitudes brighter than the median of a rolling window has by far the greatest influence in discriminating between the classes.

\subsubsection{Light curve only model}

\begin{figure*}
    \centering
    \begin{subfigure}[b]{\columnwidth}
        \centering
    	\includegraphics[width=\columnwidth]{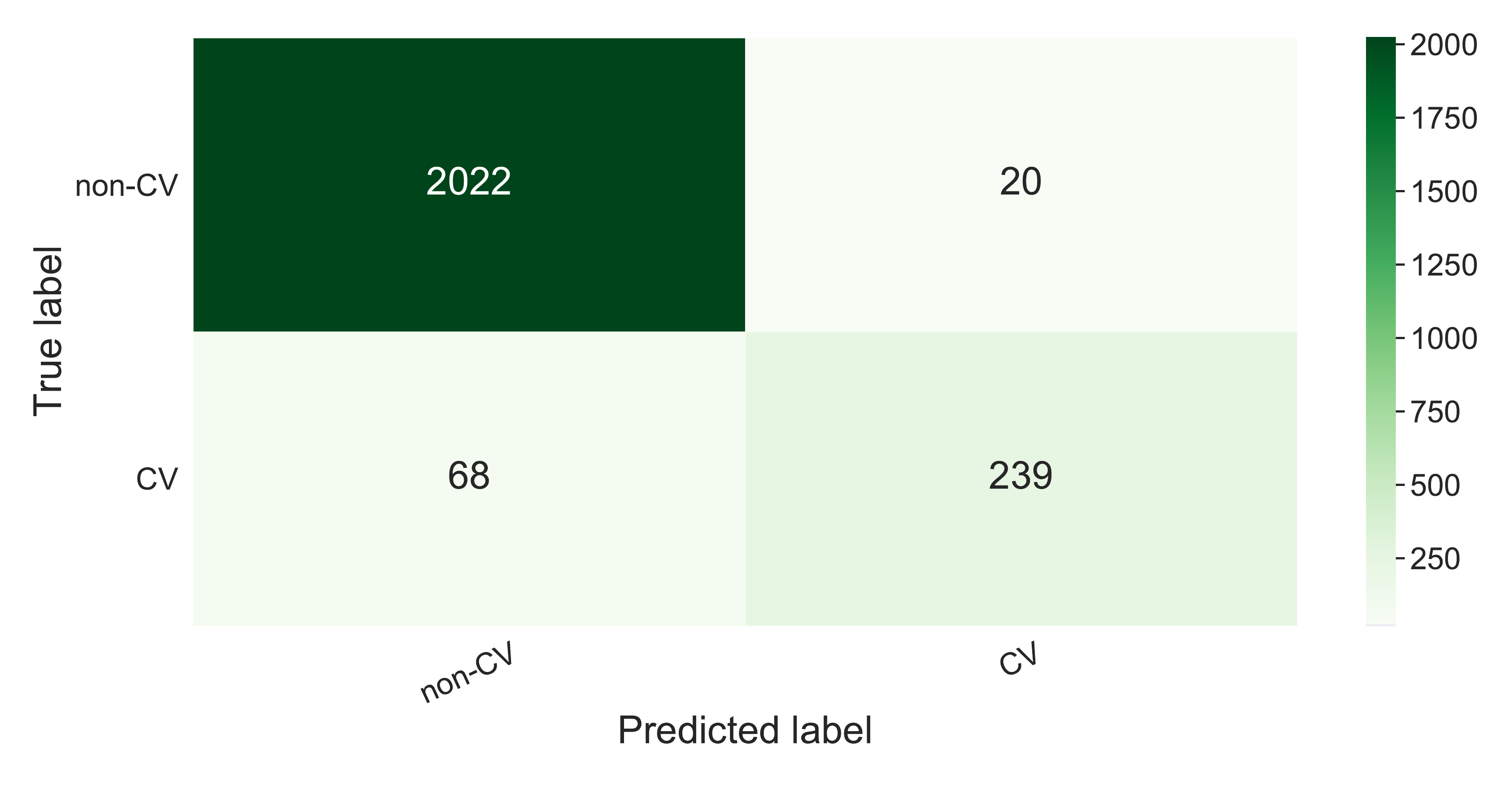}
        \label{fig:Binary_XGB_FF_CM}
    \end{subfigure}
    \hfill
    \begin{subfigure}[b]{\columnwidth}
        \centering
    	\includegraphics[width=\columnwidth]{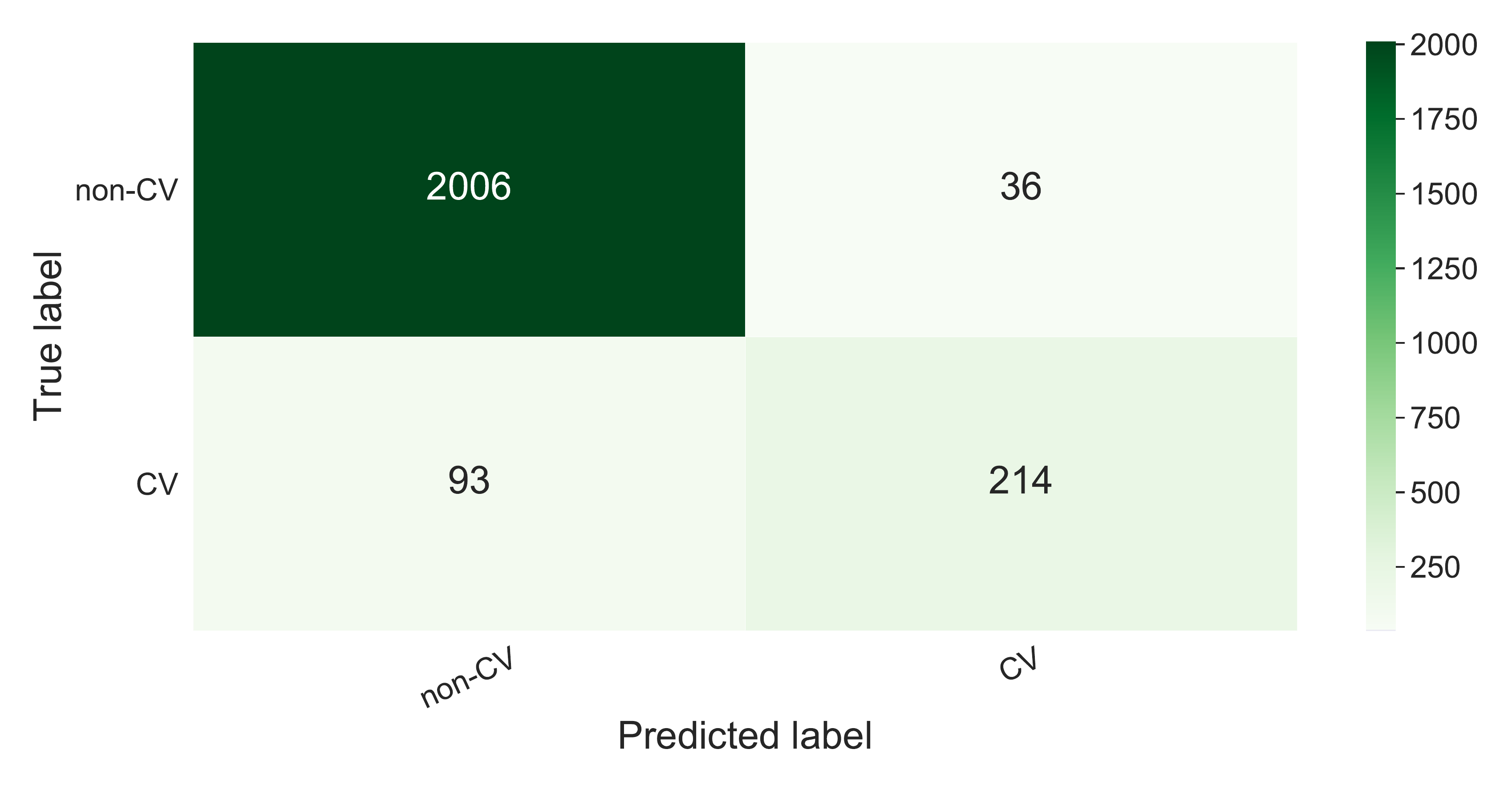}
        \label{fig:Binary_XGB_LC_CM}
    \end{subfigure}
    \caption{Confusion Matrices (CM) for the best performing binary task full feature (left) and light curve only (right) models. In each case this was an XGBoost model - achieved the highest f1-score. The CMs show the numbers corresponding to precision, recall and accuracy scores in table \ref{tab:Results_Binary}. There are over 6 and a half times more non-CVs in our test set than CVs, raising the overall accuracy score, the balanced accuracy score is more able to account for this class imbalance.}
    \label{fig:Binary_CMs}
\end{figure*}

\begin{figure*}
    \centering
    \begin{subfigure}[b]{\columnwidth}
        \centering
    	\includegraphics[width=\columnwidth]{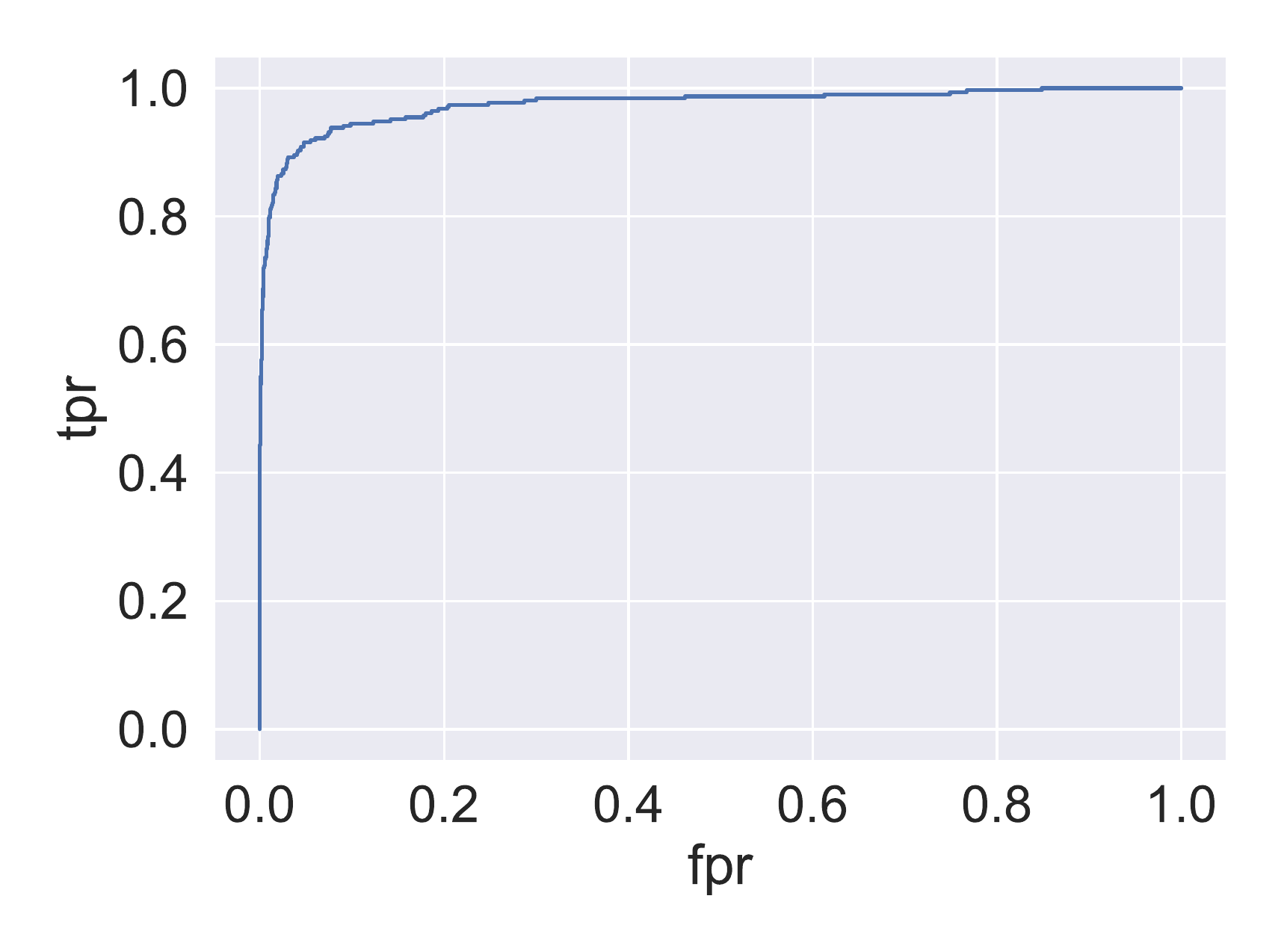}
        \label{fig:Binary_XGB_FF_AUC}
    \end{subfigure}
    \hfill
    \begin{subfigure}[b]{\columnwidth}
        \centering
    	\includegraphics[width=\columnwidth]{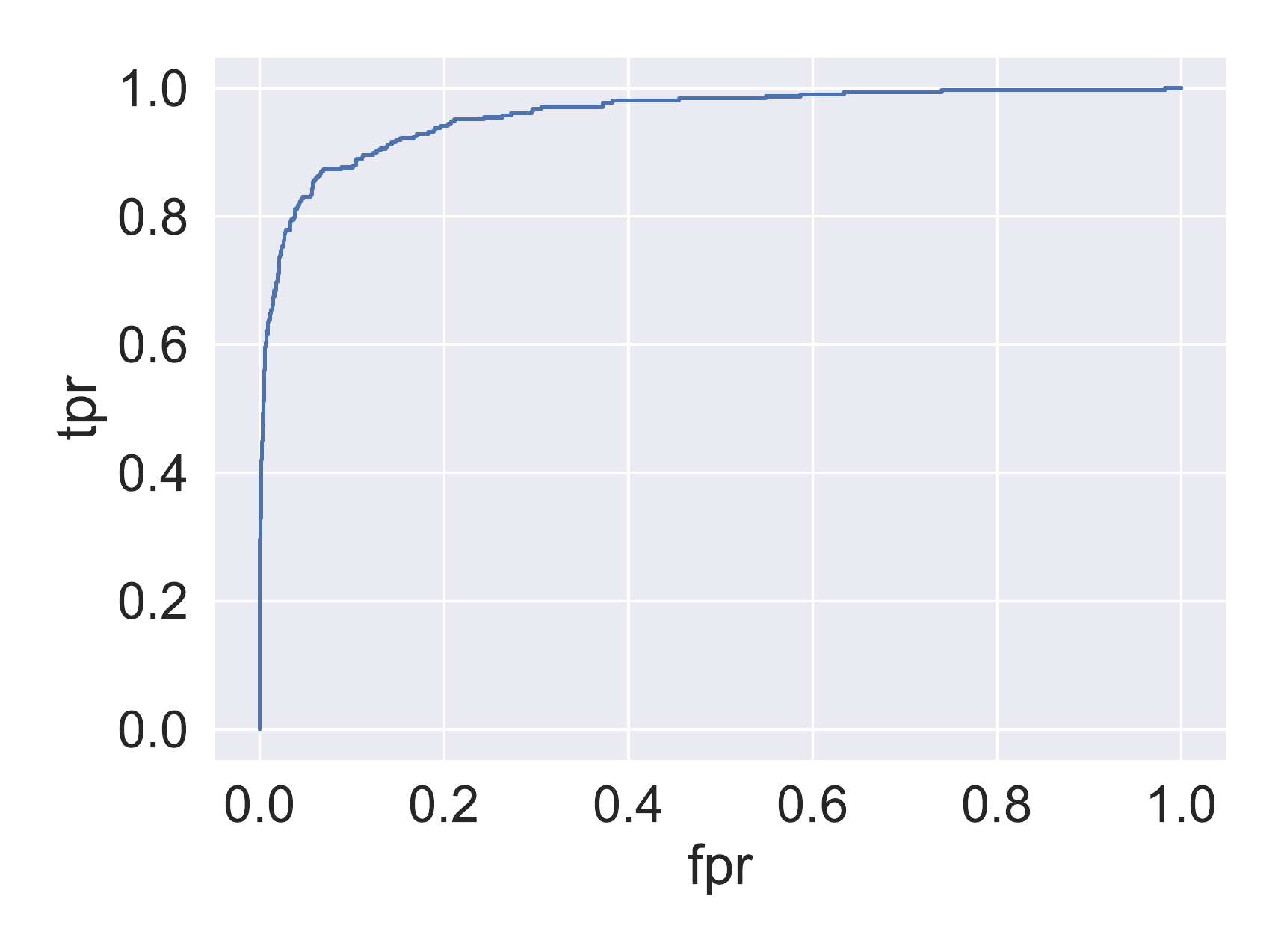}
        \label{fig:Binary_XGB_LC_AUC}
    \end{subfigure}
    \caption{ROC curves for the full feature and light curve only binary task models achieving the highest CV f1-scores. On the left is the curve for the full feature model, while on the right is that for the light curve only model. The full feature model area under the curve is 0.975, for the light curve only model this is 0.9622, indicate a strong performance in each case.}
    \label{fig:Binary_XGB_AUC}
\end{figure*}

\begin{figure*}
    \begin{subfigure}[b]{\columnwidth}
        \centering
    	\includegraphics[width=\columnwidth]{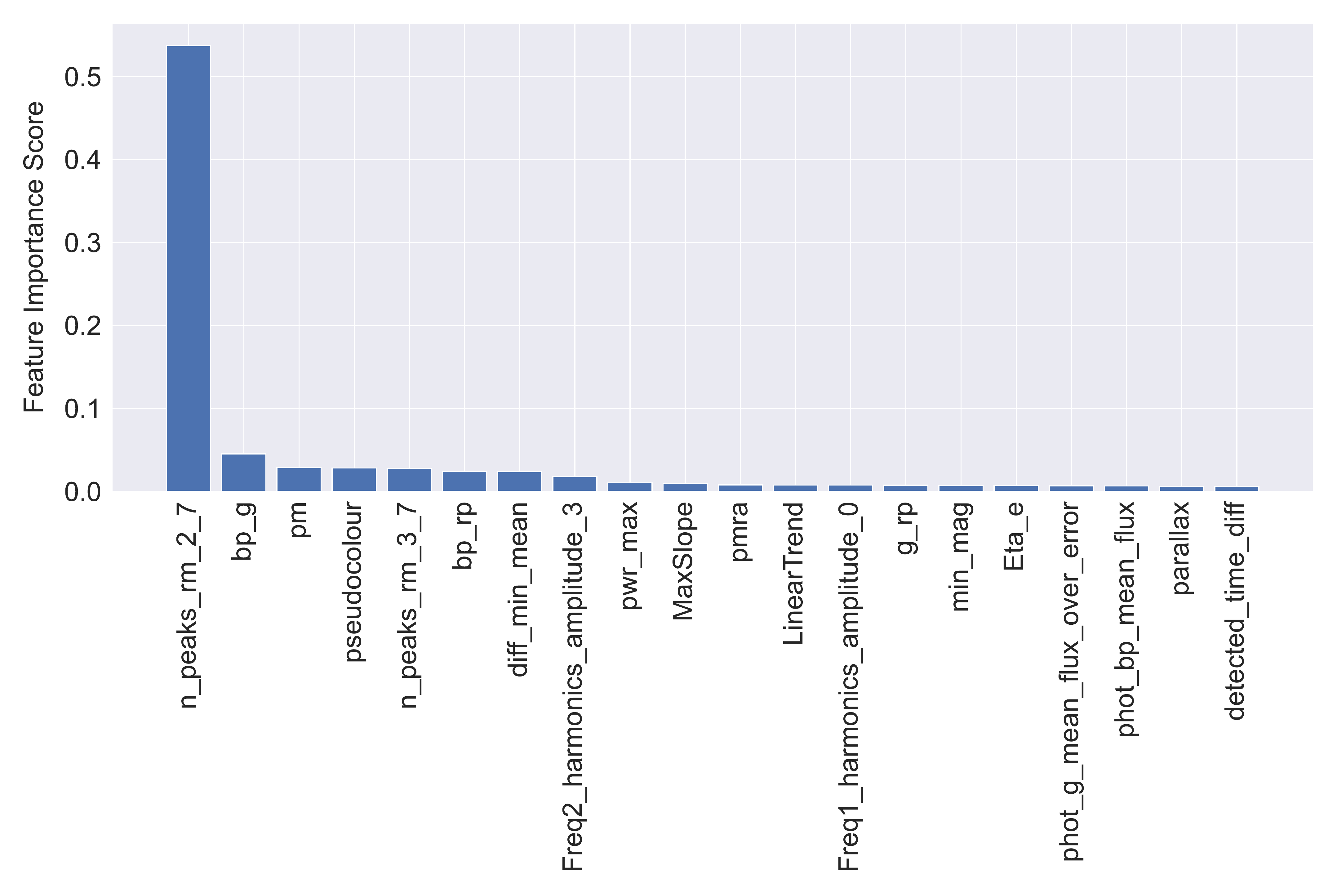}
        \label{fig:Binary_XGB_FF_FI}
    \end{subfigure}
    \hfill
    \begin{subfigure}[b]{\columnwidth}
        \centering
    	\includegraphics[width=\columnwidth]{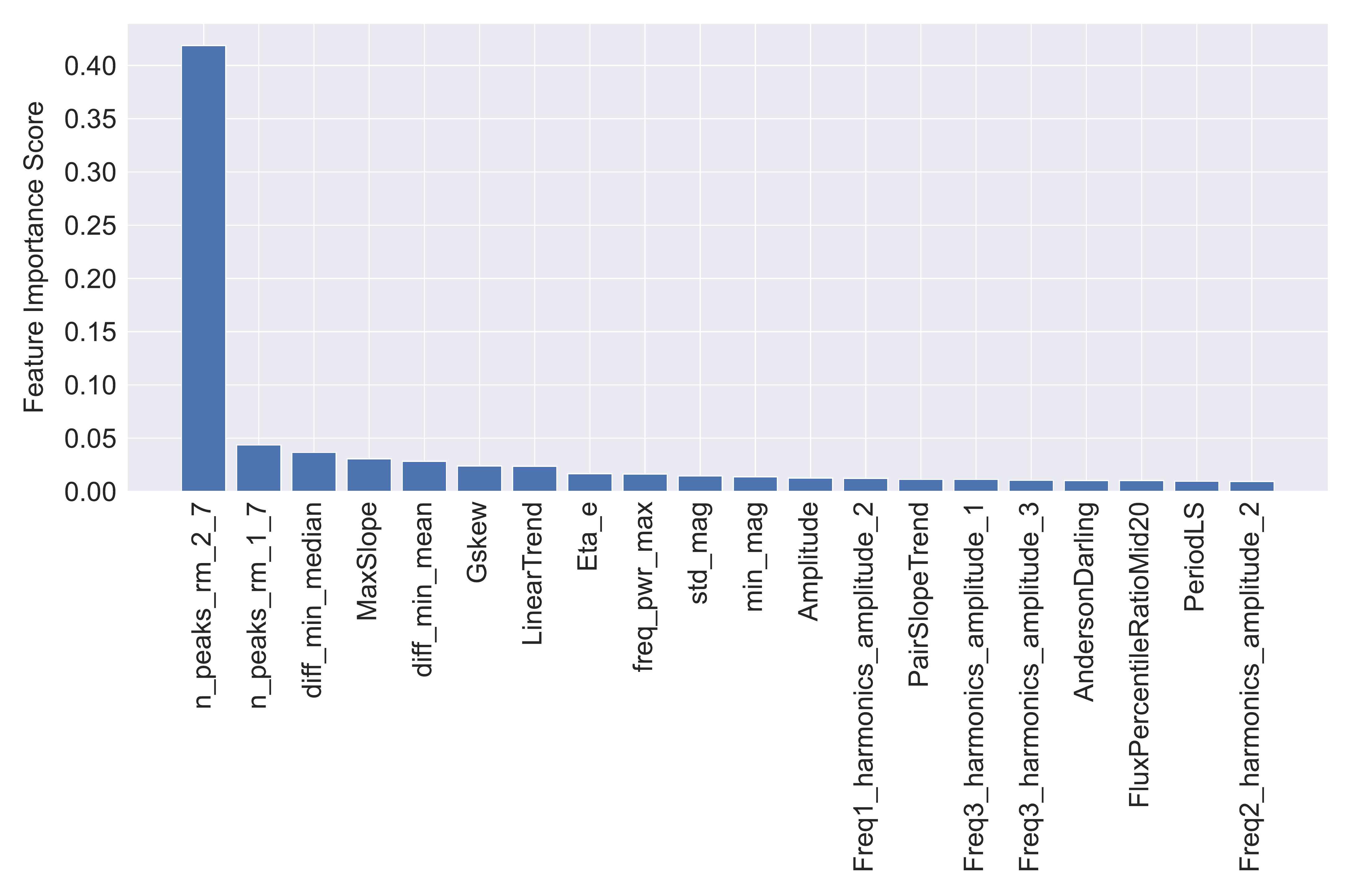}
        \label{fig:Binary_XGB_LC_FI}
    \end{subfigure}
    \caption{Feature importance scores for the 20 most influential features within the best performing full-feature and light curve only binary task models. Feature importance refers to a class of techniques for assigning scores to input features to a predictive model, in this case XGBoost, that indicates the relative importance of each feature when making a prediction. The most important feature for each of the full feature (left) and light curve only (right) models is \textit{n\_peaks\_rm\_2\_7} - number of instances of data points at least 2 magnitudes brighter than the median of a rolling window of 7 epochs. Feature definitions are contained in tables \ref{tab:Feature_w/o_feets}, \ref{tab:feets} and \ref{tab:EDR3}.}
    \label{fig:Binary_XGB_FI}
\end{figure*}

The best performing binary task light curve only model was XGBoost (CV f1-score of 76\%). The implementation was performed with 150 Decision Trees at a learning rate of 0.2 and maximum tree depth of 6. The Random Forest models follow closely behind (CV F1-score of 74\%); use of McNemar's test again showing XGBoost makes errors in the same proportions ($0.766\leq p\leq0.88$). The CV f1-score performance for this XGBoost model drops compared to the full feature model by 8 percentage points due to an increase in the number of false negatives from 69 to 93 and an increase in the number of false positives to 36 from 20. Out of the 307 test set CVs, 214 were correctly identified (see right panel of figure \ref{fig:Binary_CMs}. The model AUC score also drops from 0.975 to 0.9622 (right panel of figure \ref{fig:Binary_XGB_AUC}. The number of observations greater than 2 magnitudes brighter than the median of a rolling window remains the feature that has by far greatest influence in discriminating between the classes (right panel of figure \ref{fig:Binary_XGB_FI}).

\subsection{4 class classification}
\subsubsection{Full feature model}

A 750 tree Random Forest model performs equally well or better than its competitors in each of the performance metrics evaluated for this 4 class full feature task. The F1-score for CV classification stands at 89\% though the remaining ensemble learning models follow closely behind. The model was trained such that only 25\% of features (selected at random) could be used within each tree, with a maximum tree depth of 25. The confusion matrix (left panel of figure \ref{fig:4class_CMs}) displays the strong performance in distinguishing CVs from other classes. Those CVs that were miss-classified were mostly predicted to be of the SNe class (39/44). The left panel of figure \ref{fig:4class_RF_classProb} presents histograms of the probabilities of class assignment for this model. The vast majority of test set examples, 274 out of the 285 predicted CVs, were predicted as such with probabilities greater than 50\%. 79 of the 285 were predicted as CVs with a probability of 95\% or above. All but 3 examples predicted as YSOs are classified with 50\% probability or higher. 

According to the feature importances (left panel of figure \ref{fig:4class_RF_FI}) the temporal baseline of observations (\textit{detected\_time\_diff}) has the greatest influence in discriminating between classes. In addition to Gaia's observing strategy and their prevalence in the dataset, this can be partially explained by the properties of the majority class, supernova - they are too distant for their progenitors to be observable by Gaia, after several months they become too faint to be observable above the light from their host galaxy. Of the supplementary features, parallax and proper motion are expected to provide the greatest ability in class distinction, with the ability to distinguish extragalactic sources from those nearby. They both appear high in feature importances, as do the right ascension and declination error features. These errors are noticeably higher for SNe ($\sim$12.8 mas) than for remaining classes ($\sim$0.08-0.17 mas) attributed to the ability to measure these properties being affected by crowding (including contamination of light from the host galaxy).

\subsubsection{Light curve only model}

\begin{figure*}
	\centering
	\begin{subfigure}[b]{\columnwidth}
	    \centering
	    \includegraphics[width=\columnwidth]{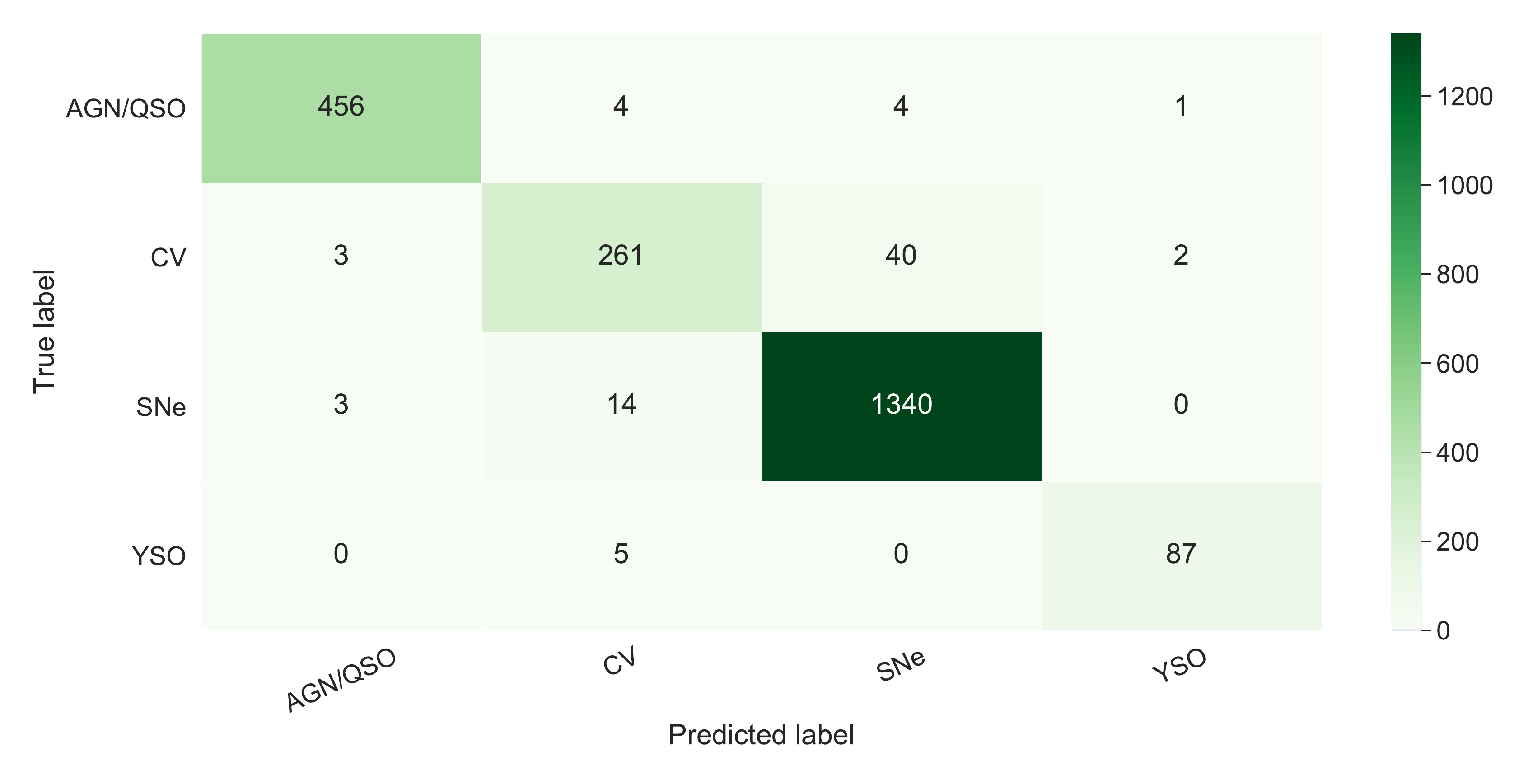}
	    \label{fig:4class_RF_FF_CM}
	\end{subfigure}
	\hfill
	\begin{subfigure}[b]{\columnwidth}
	    \centering
	    \includegraphics[width=\columnwidth]{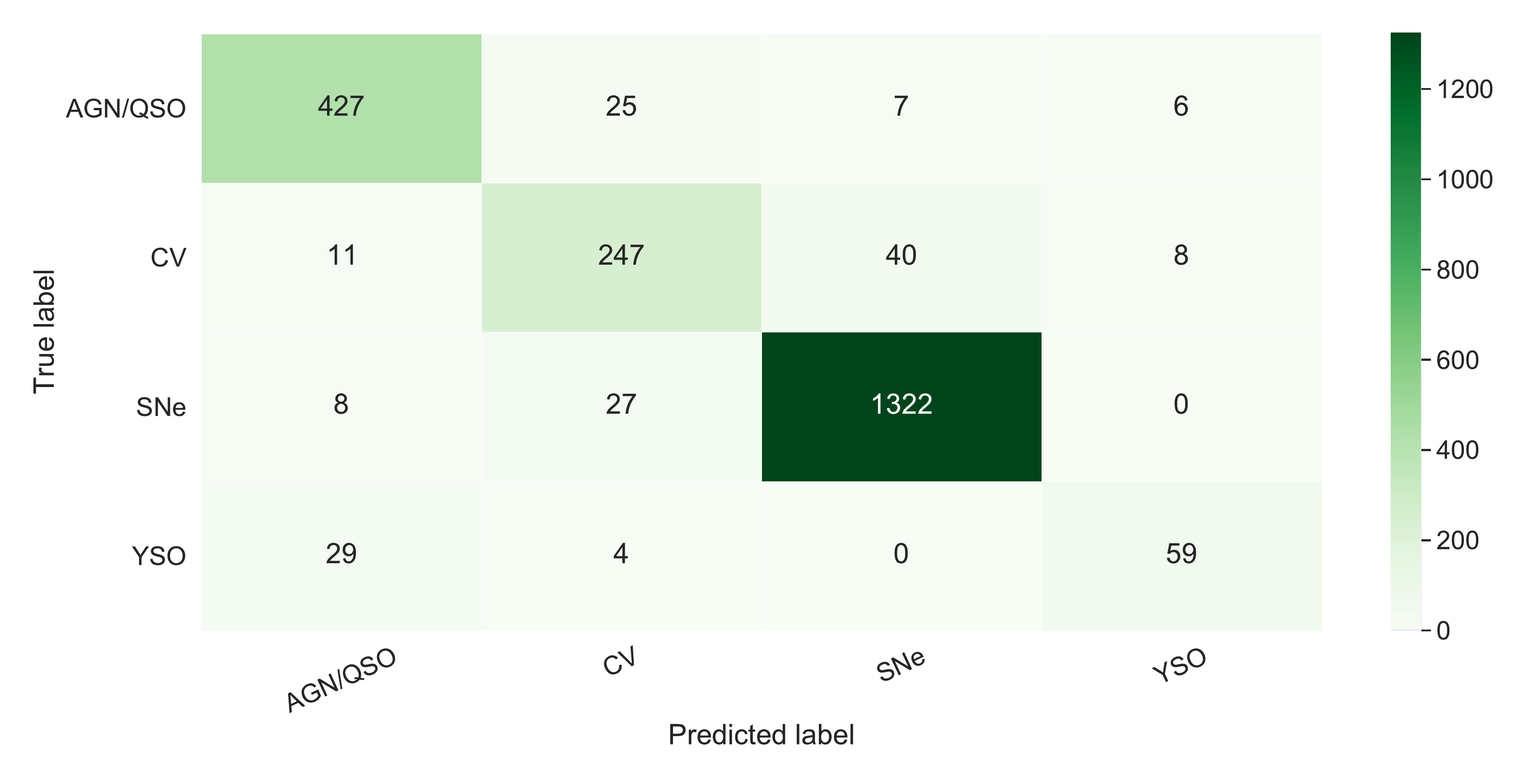}
	    \label{fig:4class_RF_LC_CM}
	\end{subfigure}
    \caption{Confusion Matrices for the best performing full feature and light curve only models in the 4 class classification task. On the left is the 750 tree Random Forest model trained with full compliment of features. 262 of the 306 CVs in the test set were successfully classified (true positives), the majority of those miss-classified, 39 of 44, were predicted to be supernovae. On the right is the 1000 tree Random Forest model trained with light curve derived features only. Less true positives (247) compared to the full feature model. Also an increase in the number of false positives from 23 to 56, of which the majority were AGN and supernovae.}
    \label{fig:4class_CMs}
\end{figure*}

\begin{figure*}
	\centering
	\begin{subfigure}[b]{\columnwidth}
	    \includegraphics[width=\columnwidth]{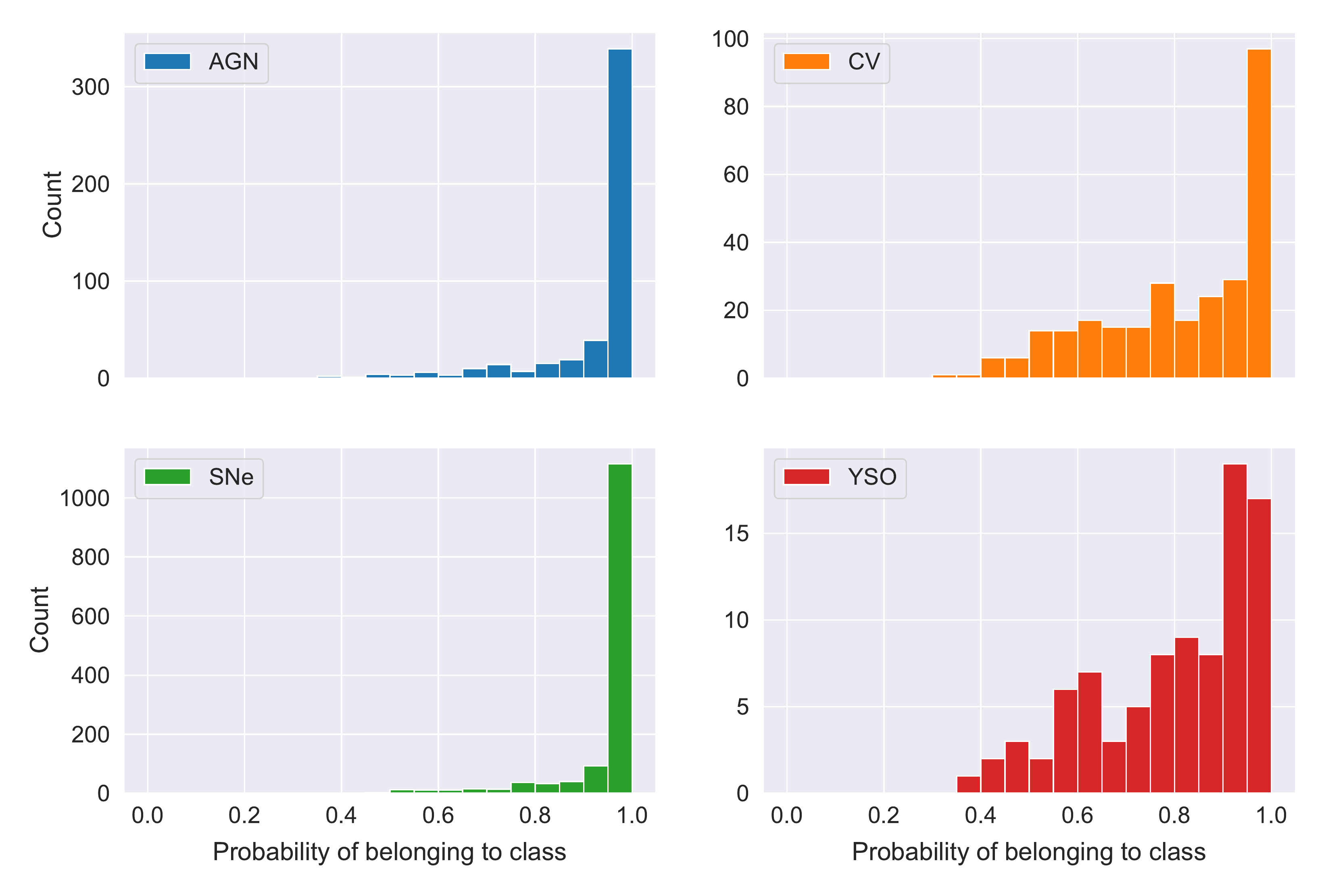}
	    \label{fig:4class_RF_FF_classProb}
	\end{subfigure}
	\rulesep
	\begin{subfigure}[b]{\columnwidth}
	    \includegraphics[width=\columnwidth]{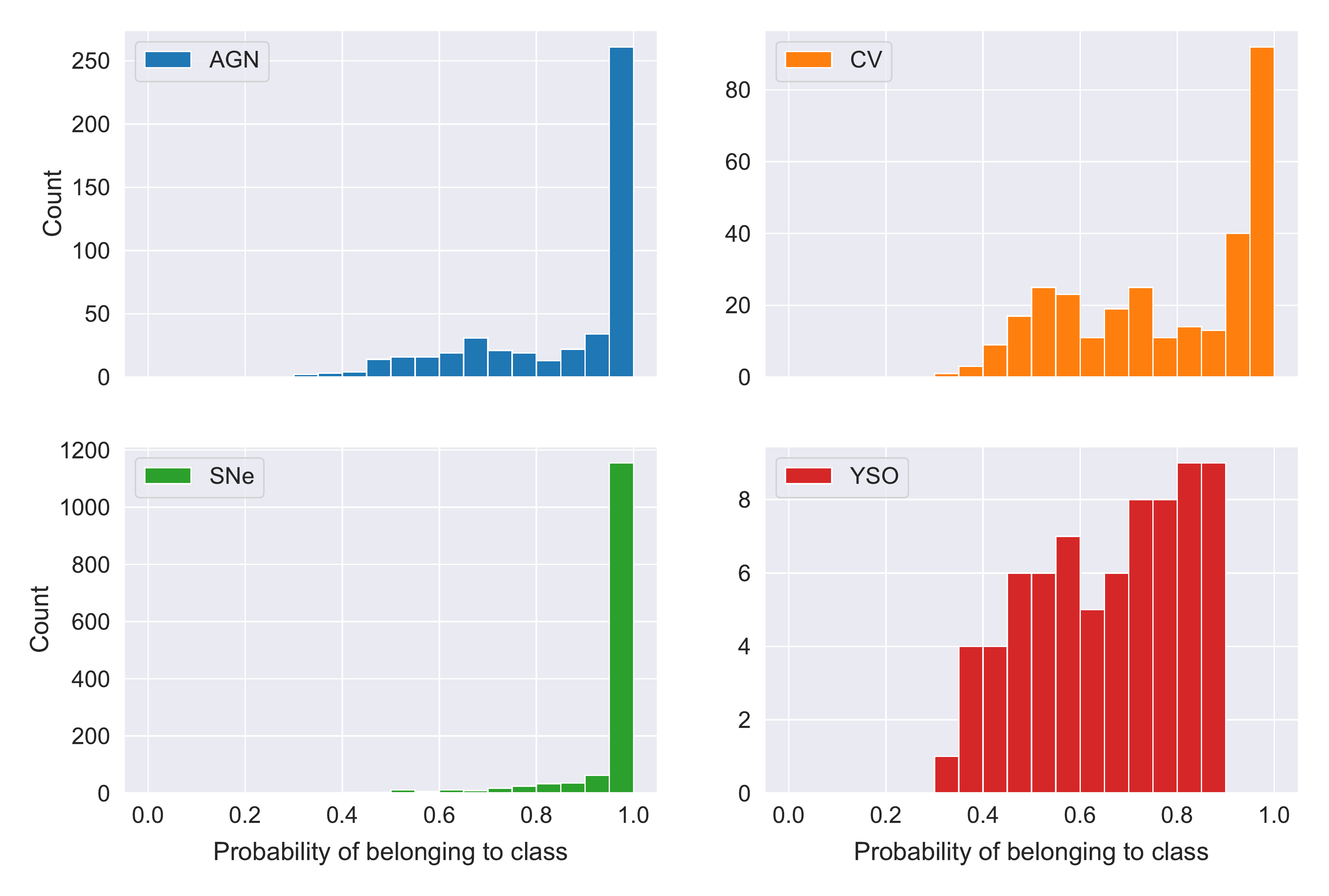}
	    \label{fig:4class_RF_LC_classProb}
	\end{subfigure}
    \caption{The number of test set examples predicted as CV (orange), AGN (blue), SNe (green), and YSO (red) separated in bins of probability of class association calculated for the full feature and light curve only 4 class models with the highest f1-scores. Each tree in the Random Forest model predicts class probabilities for each example - these are the fraction of samples of the same class in the associated leaf evaluated during training. These probabilities are averaged for the forest prediction. Class probabilities for the full feature Random Forest model (left) show nearly all examples are assigned classes with greater than 50\% probability, the majority which are in the 95-100\% bins. For the light curve only feature 4 class model (right), we can say likewise, however the YSO class assignment probabilities are more uncertain.}
    \label{fig:4class_RF_classProb}
\end{figure*}

\begin{figure*}
    \centering
    \begin{subfigure}[t]{\columnwidth }
        \centering
        \includegraphics[width=\columnwidth, height=150pt]{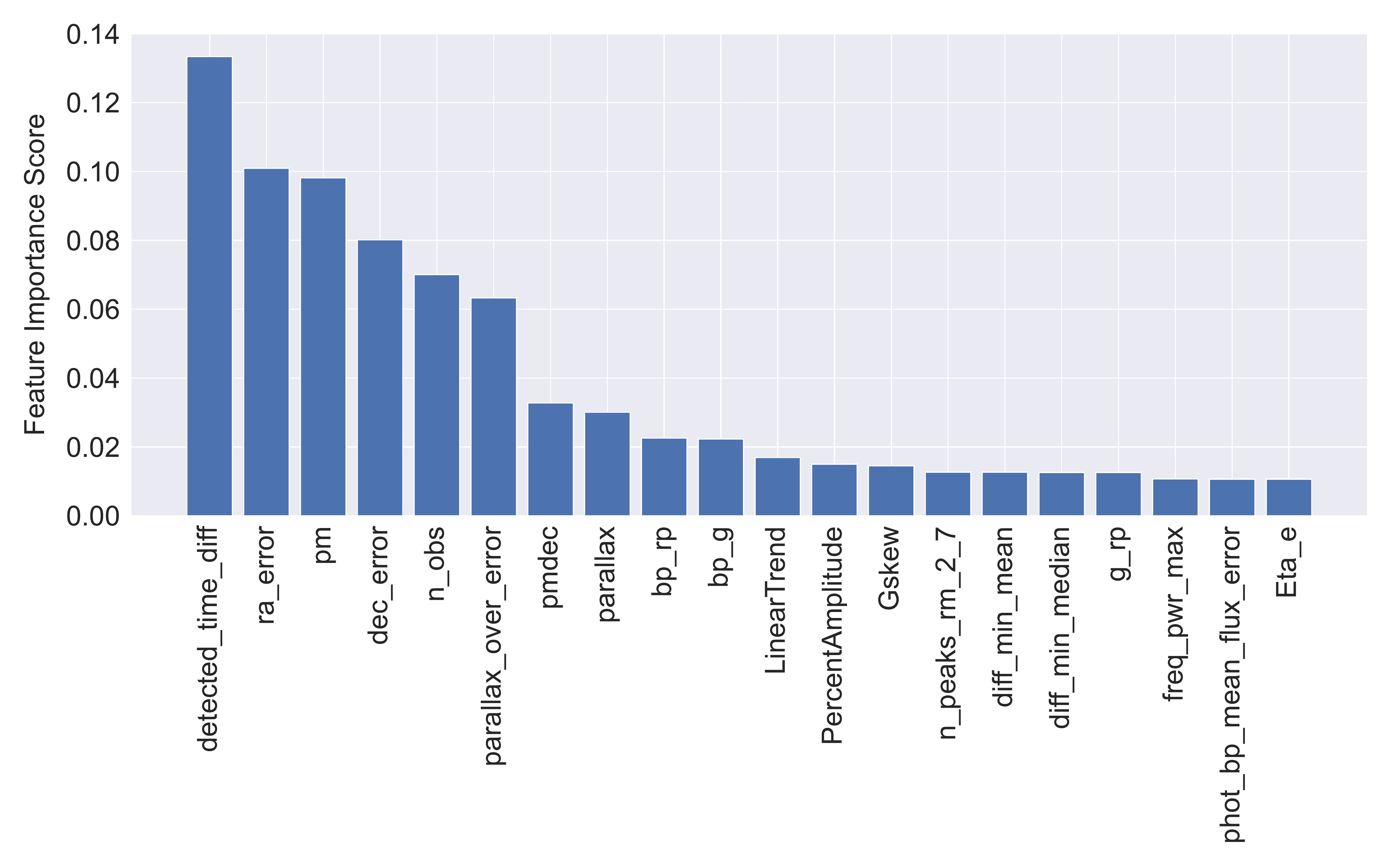}
        \label{fig:4class_RF_FF_FI}
    \end{subfigure}
    \hfill
    \begin{subfigure}[t]{\columnwidth}
        \centering
        \includegraphics[width=\columnwidth, height=150pt]{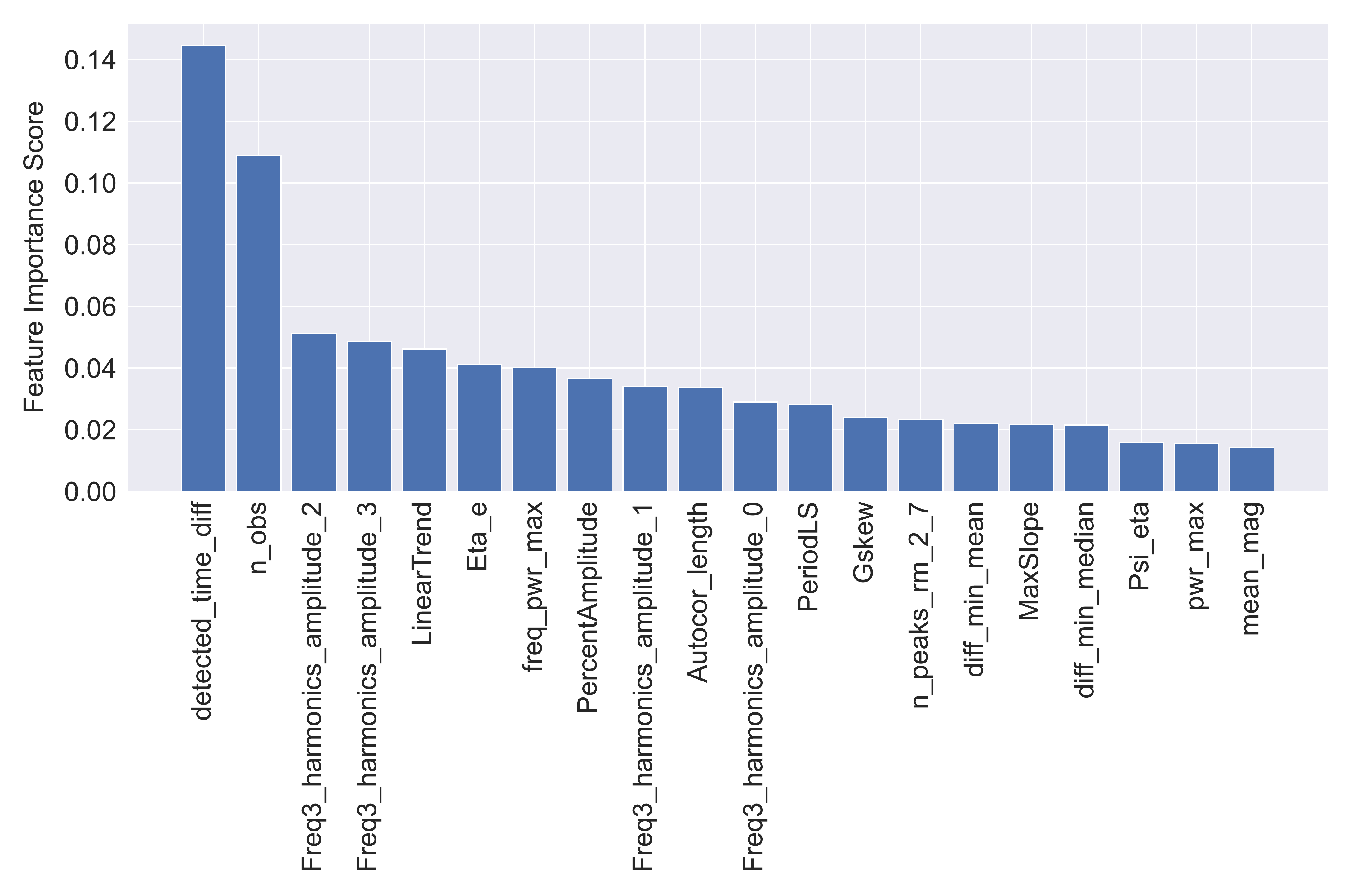}
        \label{fig:4class_RF_LC_FI}
    \end{subfigure}
    \caption{The top 20 features based on feature importance scores for the 4 class full feature and light curve only models with the highest f1-scores. The full feature model best performing feature (left) was the time between the first and last observation of the target, followed by the error in the right ascension, proper motion, and the error in the declination. The same best performing feature is present for the light curve only model (right). Feature definitions are contained in tables \ref{tab:Feature_w/o_feets}, \ref{tab:feets} and \ref{tab:EDR3}.}
    \label{fig:4class_RF_FI}
\end{figure*}

A 1000 tree Random Forest model performed the best, achieving the highest CV f1-score (81\%) for the 4 class light curve feature only task, though the remaining ensemble learning models follow closely behind. The model was implemented with a maximum tree depth of 30 and 75\% of randomly selected features available for each tree. While 247 of the 306 CVs have been correctly classified (right panel of figure \ref{fig:4class_CMs}), the contamination of other classes into those targets predicted as CV increases from 8 to 18\% compared to the full feature model. Like the full feature model, missclassified CVs are mostly assigned the SNe label. The histograms of class assignment probability (right panel of figure \ref{fig:4class_RF_classProb}) show the majority of CVs are predicted as such with greater than 50\% probability, though more examples are now present in the tail of the distribution. According to the feature importances (right panel of figure \ref{fig:4class_RF_FI}) the temporal baseline of observations (\textit{detected\_time\_diff}) also has the greatest influence in class distinction.

\section{Discussion} \label{sec:Discussion}

\subsection{Semi-regular, short duration outbursts}

The light curve feature that logs the number of epochs that are at least 2 magnitudes brighter than the median of a rolling window of 7 epochs, \textit{n\_peak\_rm\_2\_7}, outperforms all others in feature importances for the best performing full feature and light curve only binary models. The semi-regular, short duration outbursts of DNe are effectively picked out using this feature as found during its development. Such characteristics are more likely to be identified within Gaia's transient alerts pipeline than the less frequent alert triggering features of other CV subtypes so the high ranking of the feature may be expected. Indeed, a coordinate cross match with 'The Catalogue and Atlas of Cataclysmic Variables'\footnote{\url{https://archive.stsci.edu/prepds/cvcat/index.html}} \citep{RN437, RN438}, reveal 77\% of successfully cross matched dataset CVs are of listed as being DNe. Exploration of the true positives for each of the best performing binary models reveal the majority display the expected dwarf novae morphology (78 and 73\% for the full feature and light curve only models respectively).

\subsection{Limited Epoch Photometry} \label{sec:Limited Epoch Photometry}

A significant fraction of our dataset is constructed from target light curves with few epochs of observation, 36\% of targets contain 5 or fewer datapoints in their light curves. This is due to the combination of Gaia's sampling frequency and systems too faint to be observed by Gaia until a brightening event propels them into visibility. Transient phenomena more likely to display this trait will be those exhibiting a rapid and large amplitude brightening, for example SNe and the CV subclasses of classical and dwarf novae. Considering SNe comprise the majority (58\%) of our dataset, this may provide an explanation for the strong performance of \textit{detected\_time\_diff} (temporal baseline of observations) in class distinction (see figure \ref{fig:4class_RF_FI}). It may also explain the difficulty that the best performing 4 class models have in distinguishing CVs from SNe. Of the CVs missclassified by the best performing full feature 4 class model, 87\% (39 of 45) are predicted to belong to the SN class, while for the corresponding light curve only model, 68\% (40 of 59) are predicted as SNe. Similarly, the majority of missclassified SNe in each of those models are predicted to be of the CV class. Inspection of the CVs missclassified as SNe reveal the majority possess light curve morphologies that are present for the SNe samples - those with few data points (2-10 observations) and those exhibiting an approximately exponential decline with no pre-explosion data. 

\subsection{Metadata and high imputation}\label{sec:imputation}

\begin{figure*}
	\includegraphics[width=\textwidth]{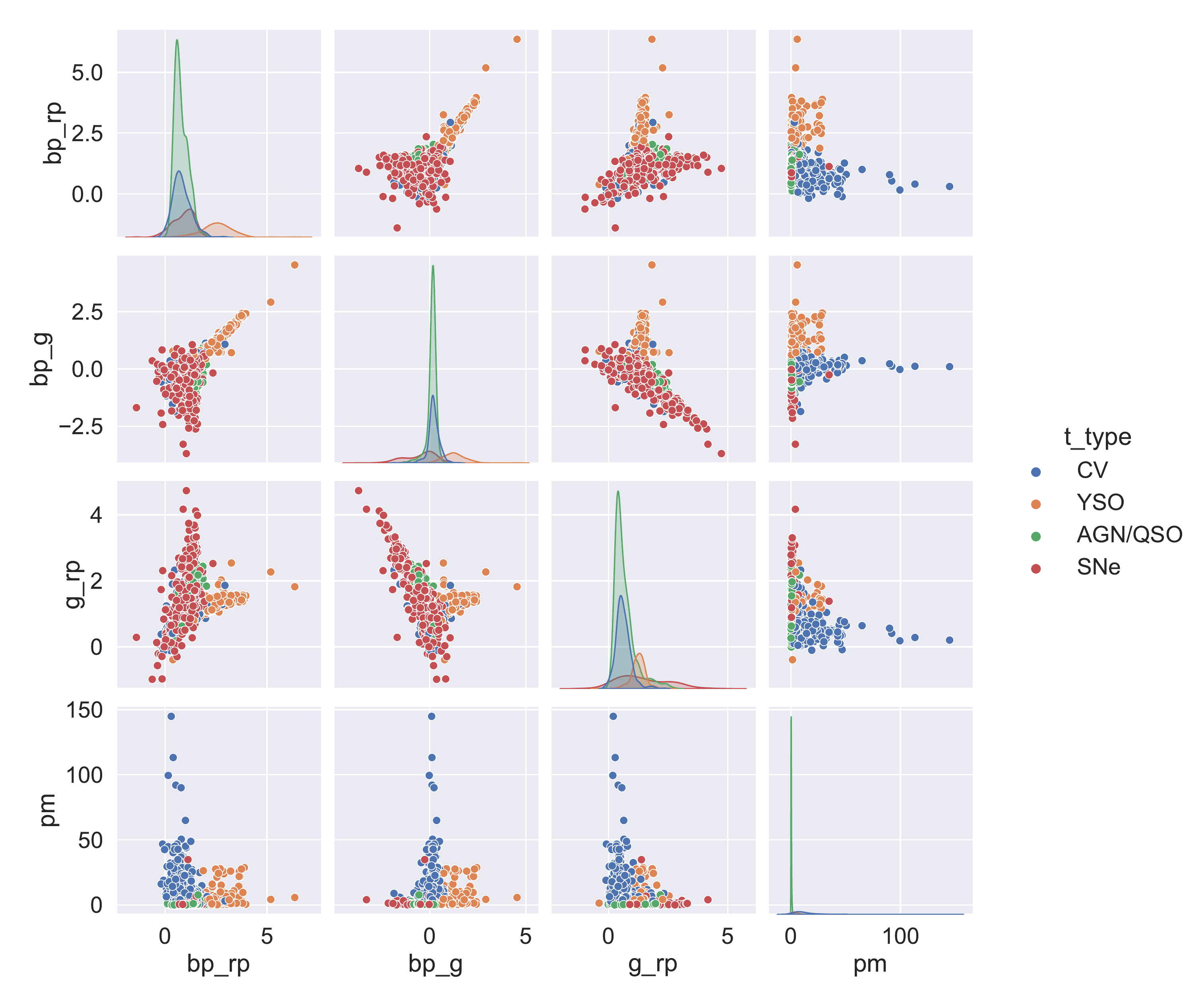}
    \caption{The pairplot allows us to see both the distribution of the single variables (plots shown diagonally from top left to bottom right) and relationships between two variables (off diagonal plots). This is shown for the bp-rp, bp-g, and g-rp colours, and proper motion. YSOs are redder in colour compared to SNe, CVs, and AGN, observed in both the single variable distribution and relationship plots, thus allowing for a significant level of class separation. Introduction of proper motion allows for separation of the more distant (extragalactic) SNe and AGN from the closer (Galactic) CV and YSO population.}
    \label{fig:PairPlot}
\end{figure*}

A McNemar's test suggests the use of metadata has an impact on model performance when comparing the full feature and light curve only XGBoost models ($p=10^{-7}$). However, the small difference in classification accuracy between these two binary models (1.9\%) indicates that the addition of survey metadata provides minimal benefit in distinguishing CVs from non-CVs. This is also shown by the small difference in the AUC (1.3\%) between these models, with both performing strongly by this measure (0.975 and 0.962 for the full feature and light curve only models respectively). The feature importances for both binary models further illustrates this point - the influence of supplementary features in class distinction is dwarfed by the light curve derived \textit{n\_peaks\_rm\_2\_7} feature. Either the metadata is unimportant or mean imputation has diluted the influence this data has on class distinction. The latter seems more likely when presented with pair plots of figure \ref{fig:PairPlot} that show transient classes in metadata feature space. This plot is of particular use in interpreting the performance of algorithms that rely on class separation within feature space (e.g., KNN and SVM). Evident is the distinction between YSOs from CVs, SNe, and AGN in colour space (bp-rp, bp-g, g-rp); and CVs and YSOs from SNe and AGN when proper motion is considered.

Use of mean imputation has its drawbacks, it ignores relationships between features, the correlation for example, and reduces the variance of the variable, thereby introducing bias to our model. Furthermore, the strategy may not be suitable for several supplementary features. For example, the parallax may not be measurable because the object is too far away (too small to measure); and a missing value for proper motion can either be due to the object having no proper motion to measure or be due to it being too distant to be measured. A more appropriate strategy could be to replace these with a value of zero - a more accurate quantity for the parallax and proper motion of the most distant sources - though this does not account for the unavailability of these features due to an unsuccessful cross match with EDR3. While alternative methods of handling missing data could be employed (such as those summarised in \citet{RN469}), a large amount of data is missing for the supplementary features (58-90\%), this can limit the effectiveness of any such strategy \citep{RN471}. Figure \ref{fig:PairPlot} shows how photometric colour information can be an important property for class distinction, this is readily available in multi-band surveys such as ZTF and can be used to help alleviate the issue.

\subsection{Comparison with other work}

The results of our investigation compare favourably with similar classification attempts where CVs are included as class. \citet{RN65} experiments with CRTS light curves in their 8 class classification model yielded an F1-score of 75\% for the CV class, while this work exceeds this in both the binary (76\%) and 4 class (80\%) tasks where only light curve features are used. \citet{RN420} evaluated 3 different algorithms in their tiered classification attempts to distinguish between CVs, SNe subclasses, AGN, YSOs and variable star subclasses from a dataset constructed from ZTF light curves and colours from ALLWISE. Their CV recall scores for their implementation of the Balanced Random Forest \citep{RN453}, XGBoost, and Multi-Layer Perceptron classifiers are 68\%, 72\%, and 61\% respectively. This compares with 67\% and 80\% for our light curve only best performing binary and 4 class models respectively. These comparisons do not however take into account differences in the instruments used to collect the data, which translates to the nature of photometric data (e.g., observing cadence, waveband). Furthermore comparisons do not consider differences in transient classes to classify and ML methods employed.

\subsection{Gaia Unknowns}

\subsubsection{Model predictions on unknown sample}

The model that produced the highest CV f1-score overall - full feature 4 class model (Random Forest with 750 trees) - is used to make class predictions of targets labelled as `unknown' (unclassified) within the Gaia alerts stream. As of December 2021, 13,241 targets were of 'unknown' class. Of these, the model predicted 2833 (21\%) to be of the CV class, 1928, 6611, and 1869 were classified as AGN/QSO, SNe, and YSOs respectively (table \ref{tab:Predicted_CVs} contains a truncated list of targets predicted as CV). As mentioned in Section \ref{sec:Dataset}, the unknown sample will contain several minority classes (e.g., microlensing and tidal disruption events) not included in the test set used to evaluate model performance. We aim to assess the impact this has on our model's ability to generalise to the unknown sample, and new transient alerts in general. This will require spectroscopic observations for a statistically significant fraction of our 2833 predicted CVs to identify their true transient classification.

\subsubsection{Spectroscopic follow-up}

We are therefore undertaking a pilot study to assess the performance of our model and the methods used by obtaining spectroscopic observations to classify those targets that can be observed with the SPRAT low resolution spectrograph \citep{RN409} mounted on the Liverpool Telescope (LT; \citealt{RN455}). These spectra cover a wavelength range of 4000 to 8000\AA \space with a resolution of 18\AA, corresponding to a resolving power R=350 at the centre this range. A limit on telescope time and the need for high quality spectra requires of an efficient observation strategy. Accordingly, observations are limited to targets with a median brightness no fainter than 18th magnitude. Furthermore, only those targets that rise highest in the sky - visible for longer at a lower airmass- are considered. Therefore we limit our sample to those with a declination corresponding to an altitude no lower than 50 degrees when at transit altitude. These cuts leave a sample of 220 targets, 7.8\% of the total catalogue, representing a statistically significant fraction with which we can validate the performance of our model. We have spectroscopically classified 15 of this sample, all of which we can confirm are of the CV class. Details of these targets are given in Table \ref{tab:CV_SPRAT}, while the associated SPRAT spectra are shown in Figure \ref{fig:SPRAT_Spectra}. Classification as a CV is based on the presence of Balmer and/or \ion{He}{I}/\ion{He}{II} lines. Where the signal to noise ratio of the spectrum permits, subtype classification is performed. Full details of the spectral features we used for classification are given in \citet{RN444} and \citet{RN462}.

\begin{figure*}
    \centering
    \begin{subfigure}[b]{0.94\columnwidth}
        \centering
        \includegraphics[width=\columnwidth]{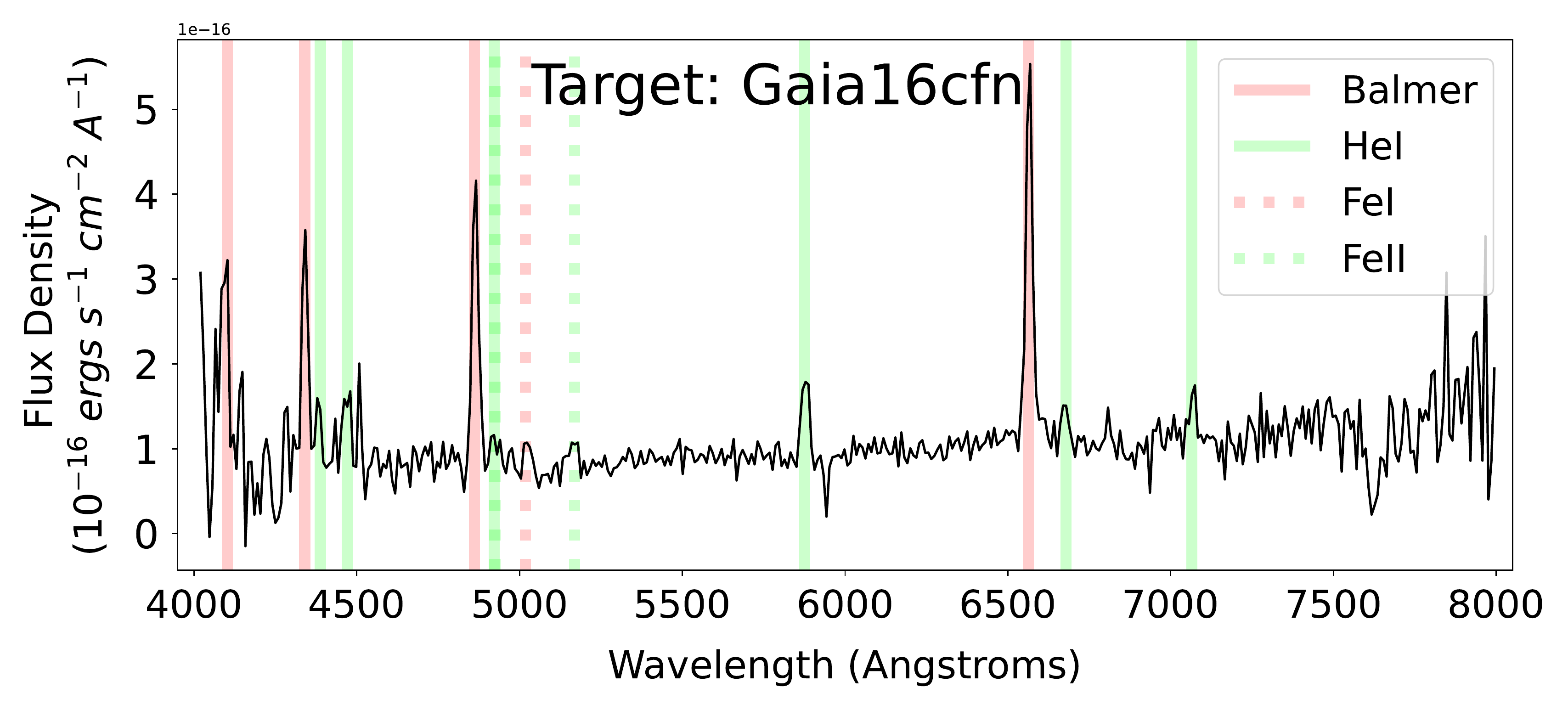}
        \label{fig:Gaia16cfn}
    \end{subfigure}
    \hfill
    \begin{subfigure}[b]{0.94\columnwidth}
        \centering
        \includegraphics[width=\columnwidth]{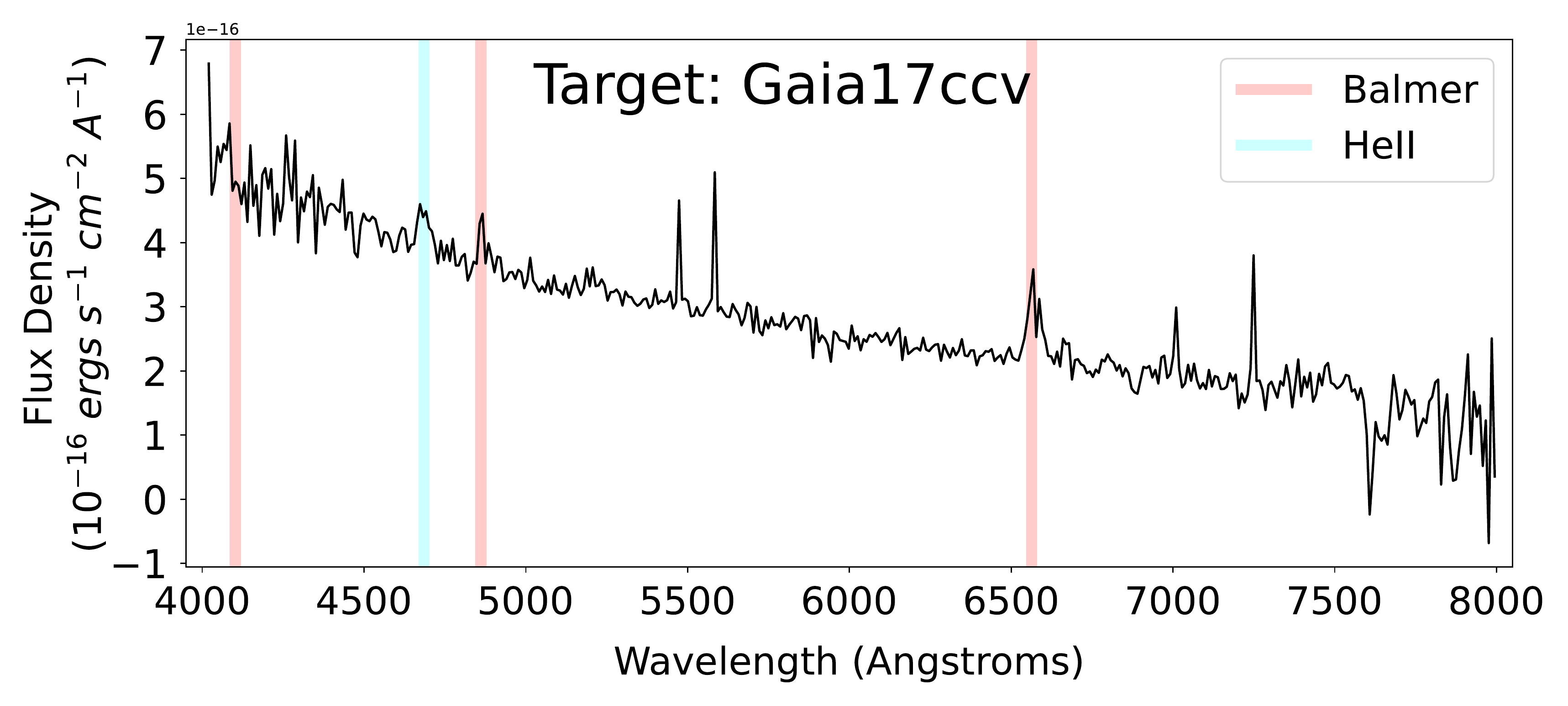}
        \label{fig:Gaia17ccv}
    \end{subfigure}
    \hfill
    \begin{subfigure}[b]{0.94\columnwidth}
        \centering
        \includegraphics[width=\columnwidth]{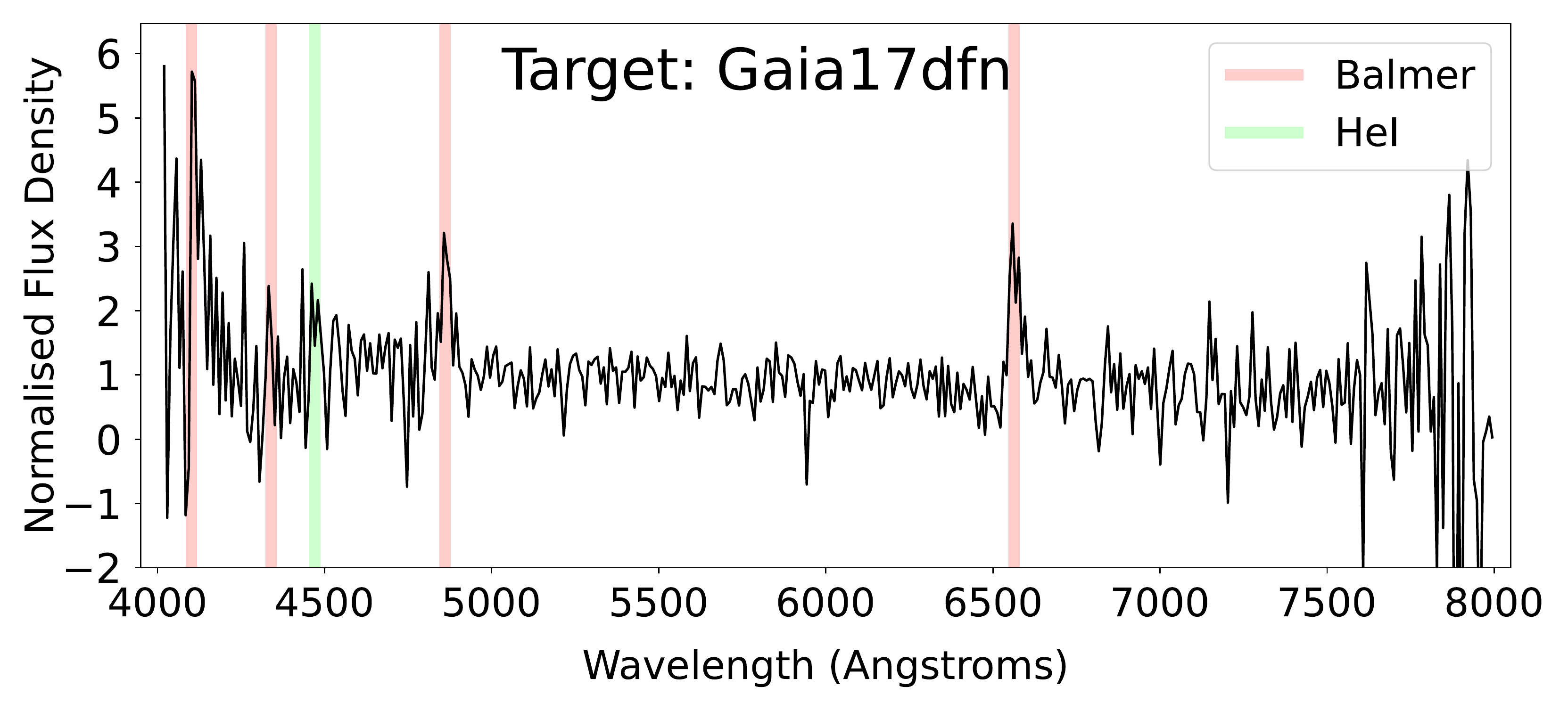}
        \label{fig:Gaia17dfn}
    \end{subfigure}
    \hfill
    \begin{subfigure}[b]{0.94\columnwidth}
        \centering
        \includegraphics[width=\columnwidth]{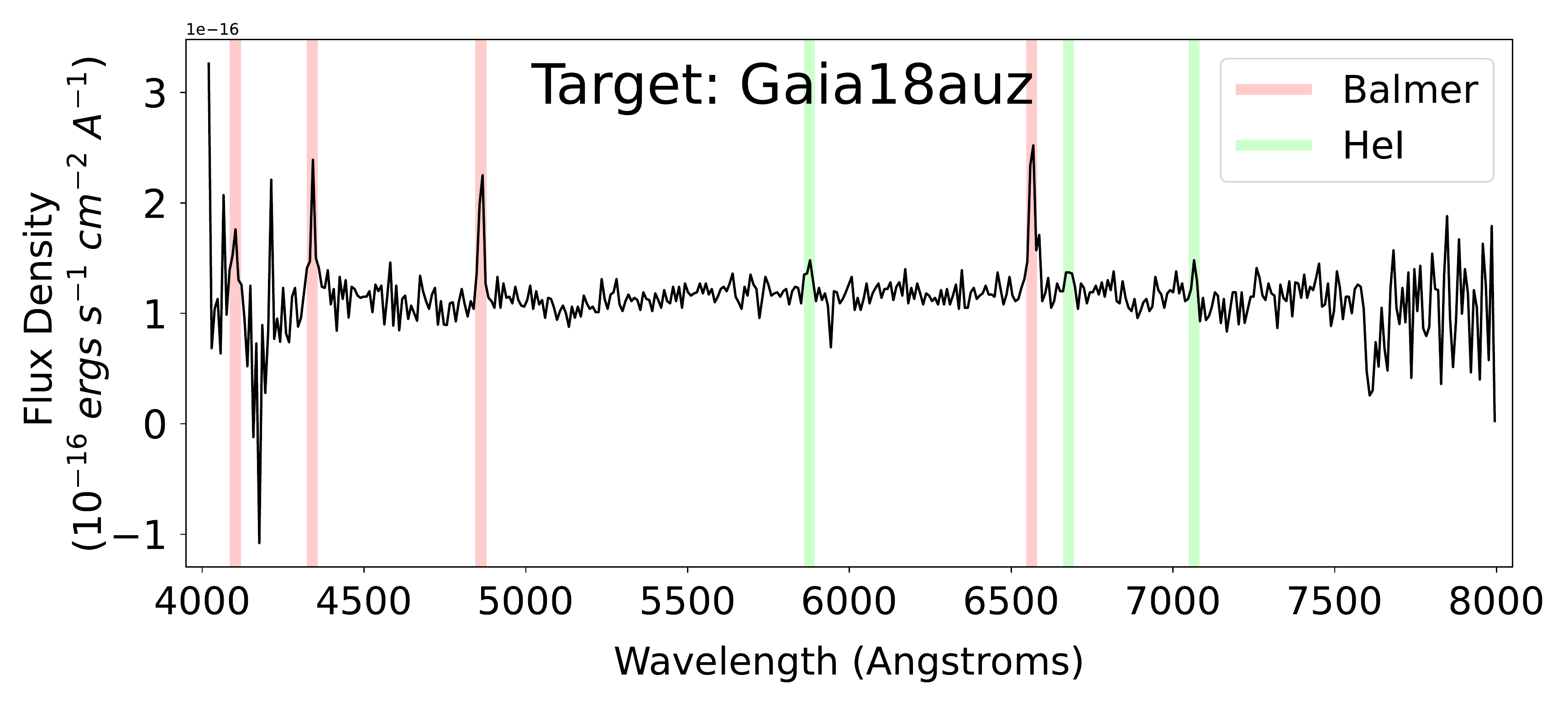}
        \label{fig:Gaia18auz}
    \end{subfigure}
    \hfill
    \begin{subfigure}[b]{0.94\columnwidth}
        \centering
        \includegraphics[width=\columnwidth]{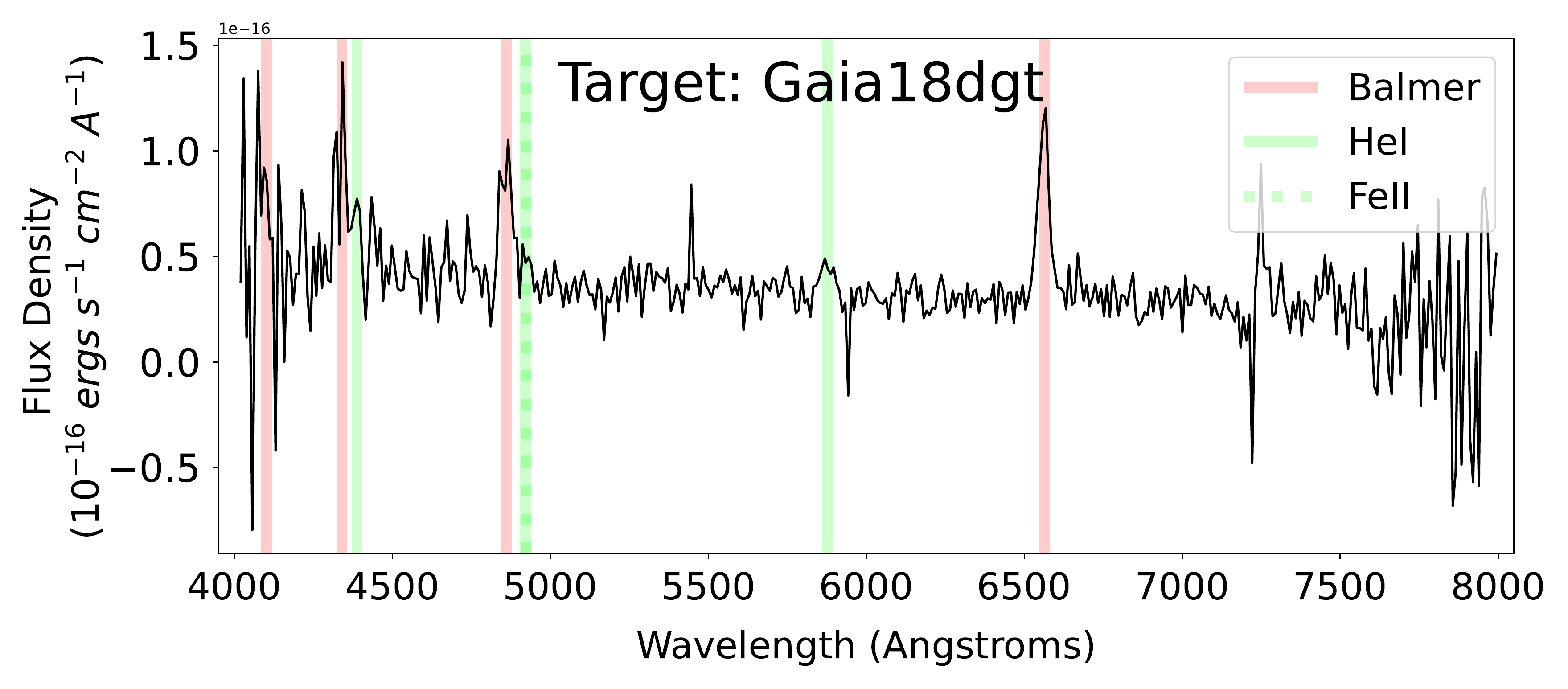}
        \label{fig:Gaia18dgt}
    \end{subfigure}
    \hfill
    \begin{subfigure}[b]{0.94\columnwidth}
        \centering
        \includegraphics[width=\columnwidth]{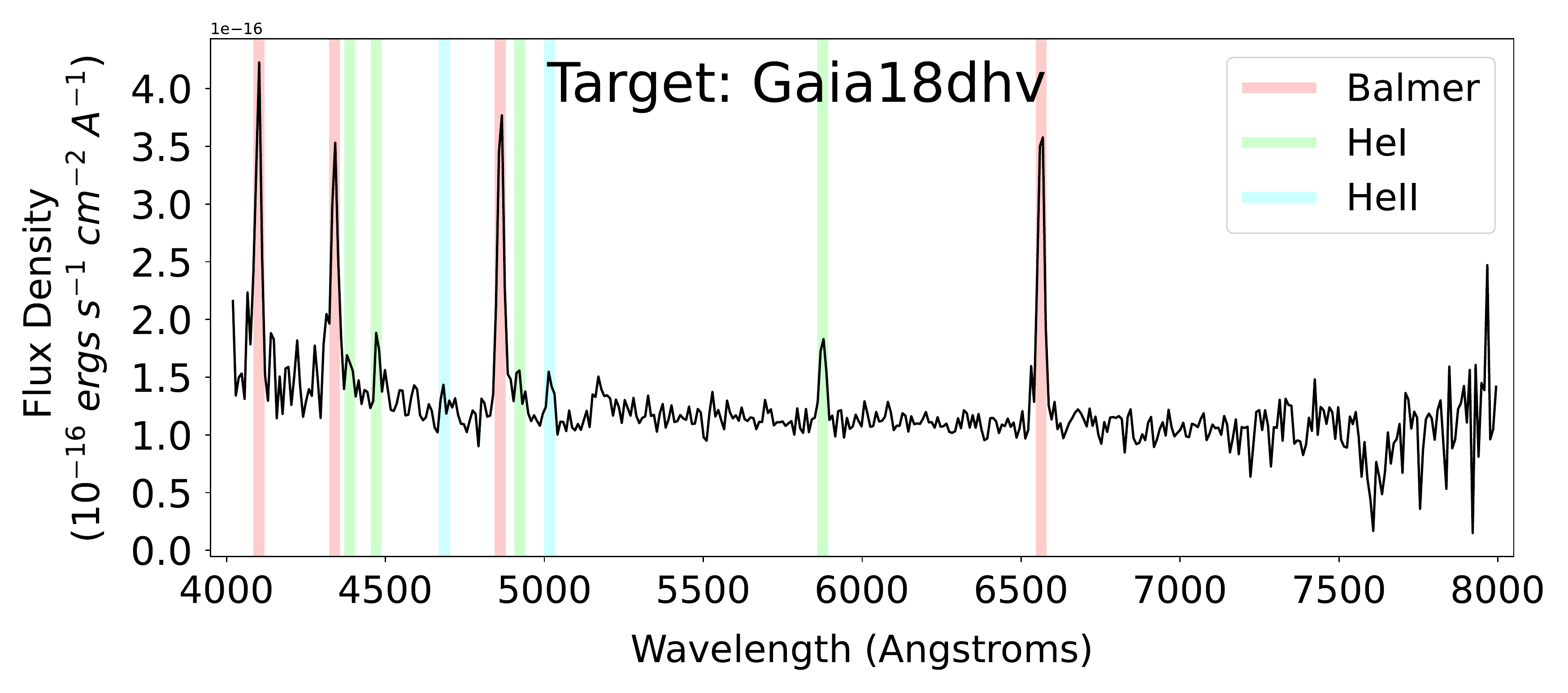}
        \label{fig:Gaia18dhv}
    \end{subfigure}
    \hfill
    \begin{subfigure}[b]{0.94\columnwidth}
        \centering
        \includegraphics[width=\columnwidth]{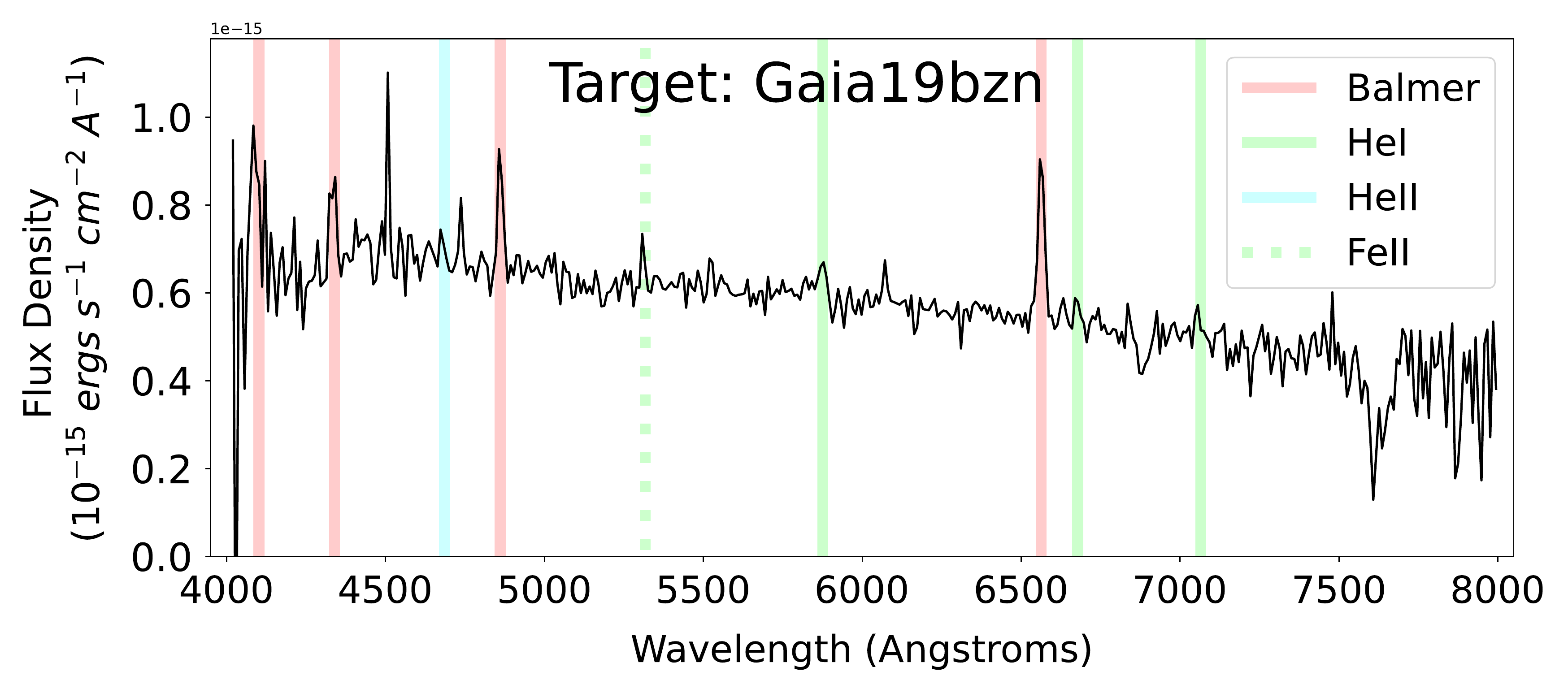}
        \label{fig:Gaia19bzn}
    \end{subfigure}
    \hfill
    \begin{subfigure}[b]{0.94\columnwidth}
        \centering
        \includegraphics[width=\columnwidth]{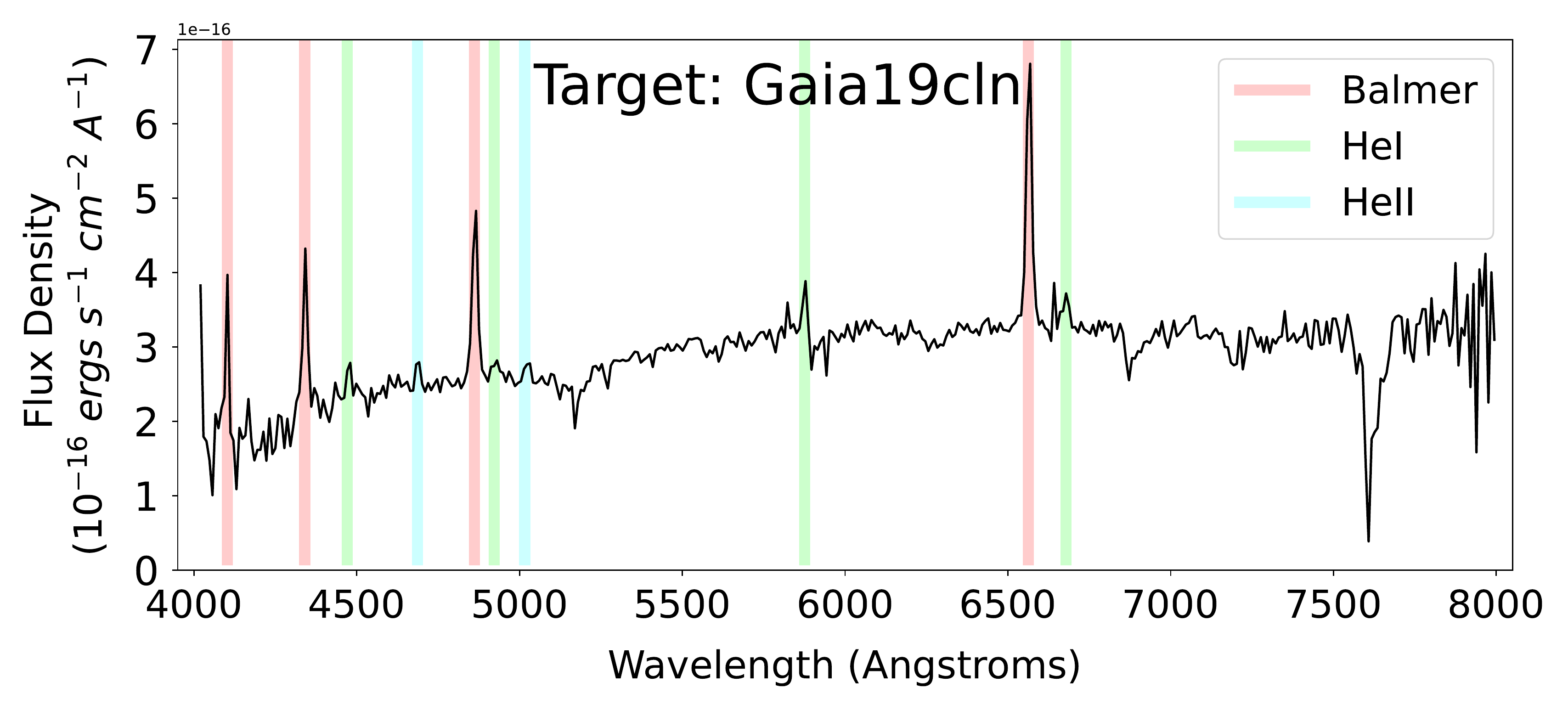}
        \label{fig:Gaia19cln}
    \end{subfigure}
    \hfill
    \begin{subfigure}[b]{0.94\columnwidth}
        \centering
        \includegraphics[width=\columnwidth]{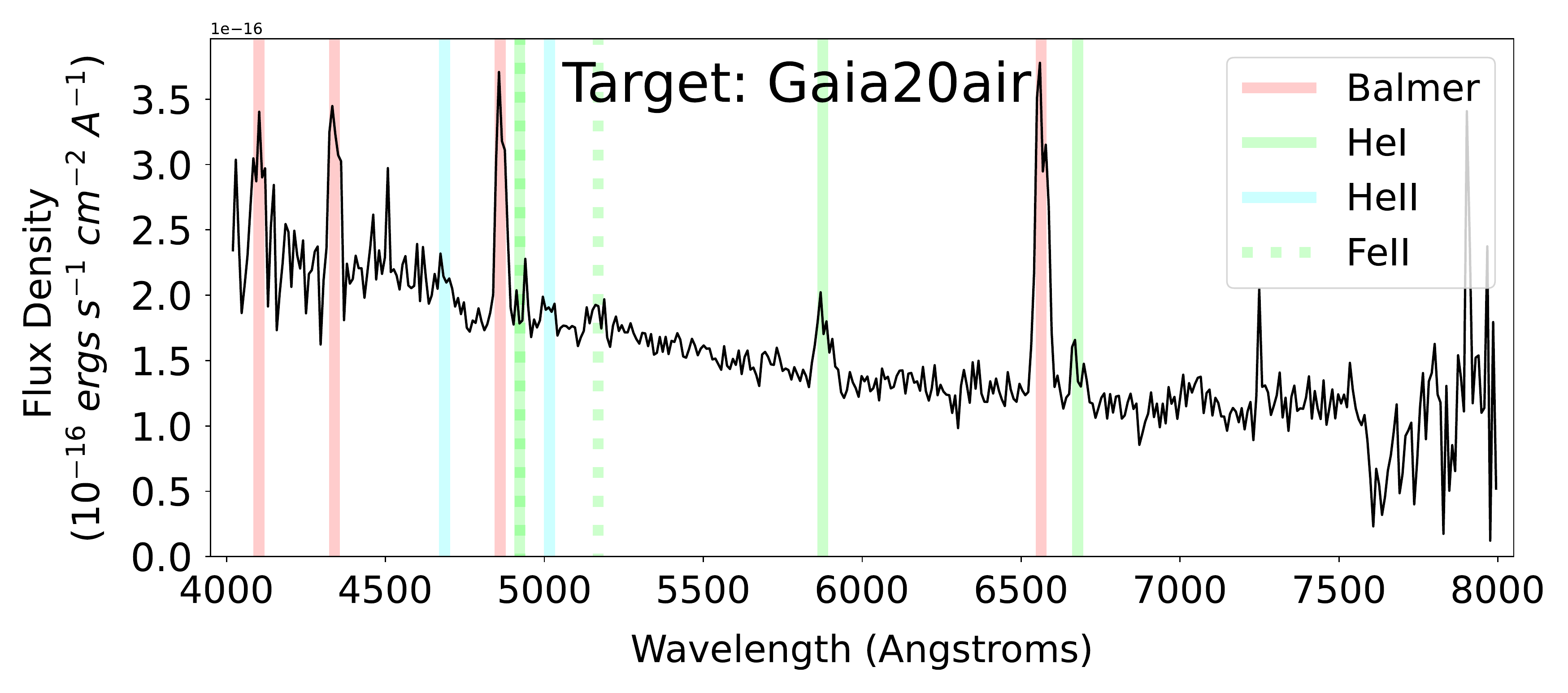}
        \label{fig:Gaia20air}
    \end{subfigure}
    \hfill
    \begin{subfigure}[b]{0.94\columnwidth}
        \centering
        \includegraphics[width=\columnwidth]{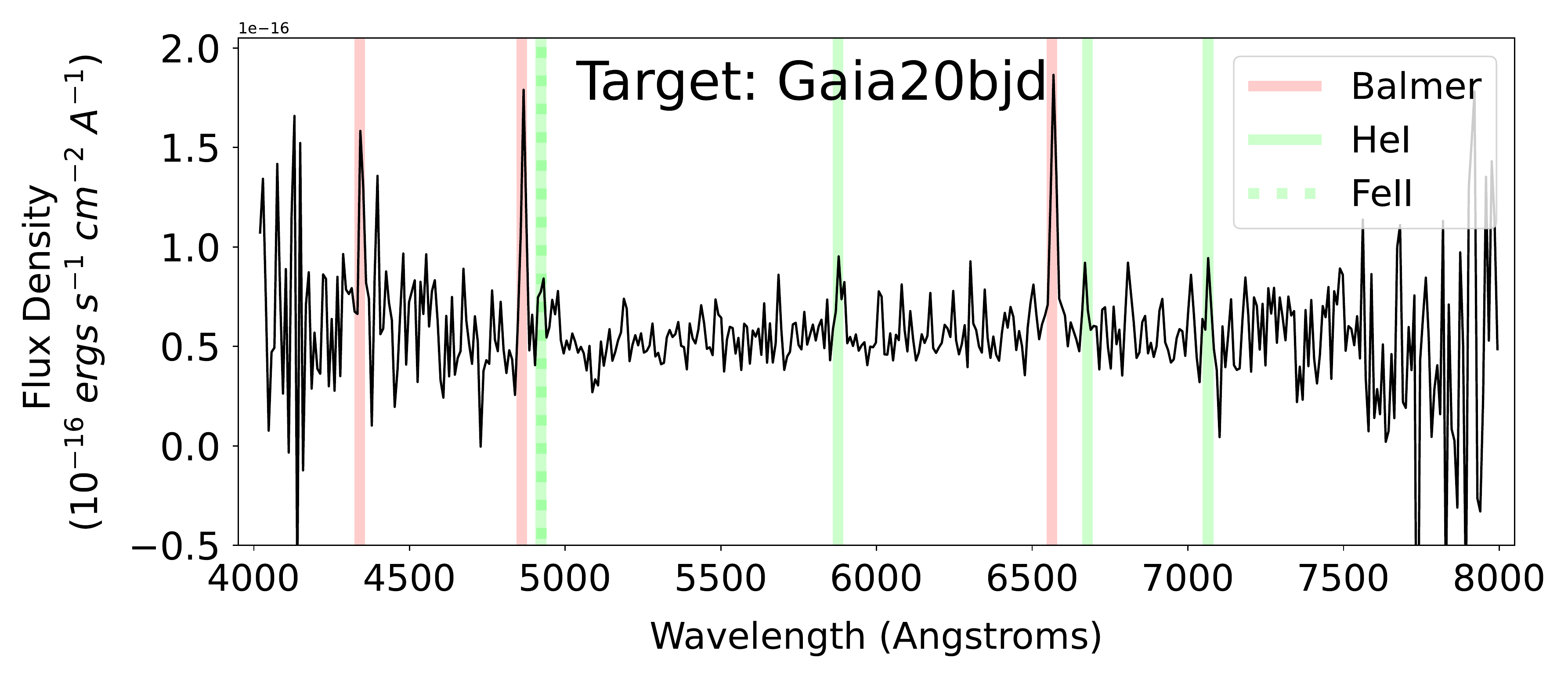}
        \label{fig:Gaia20bjd}
    \end{subfigure}
    \hfill
    \begin{subfigure}[b]{0.94\columnwidth}
        \centering
        \includegraphics[width=\columnwidth]{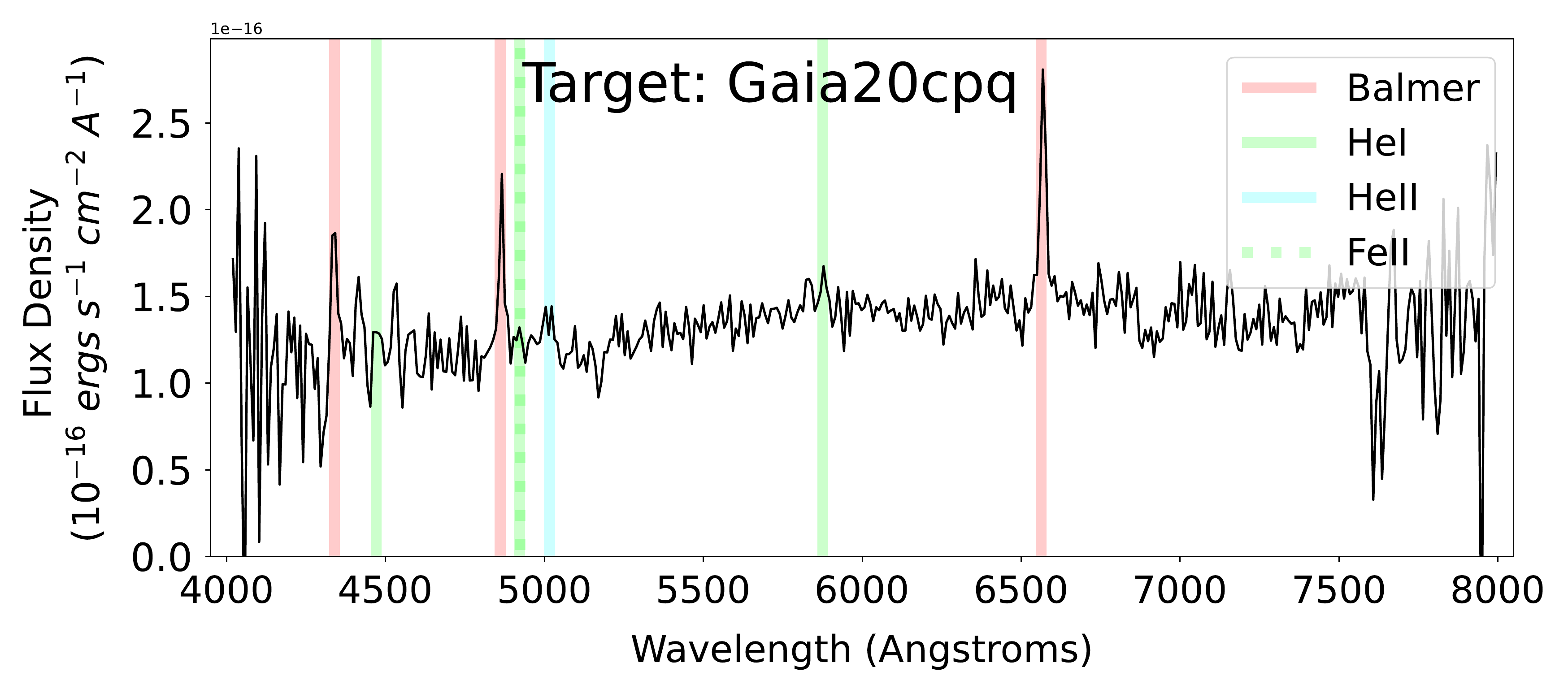}
        \label{fig:Gaia20cpq}
    \end{subfigure}
    \hfill
    \begin{subfigure}[b]{0.94\columnwidth}
        \centering
        \includegraphics[width=\columnwidth]{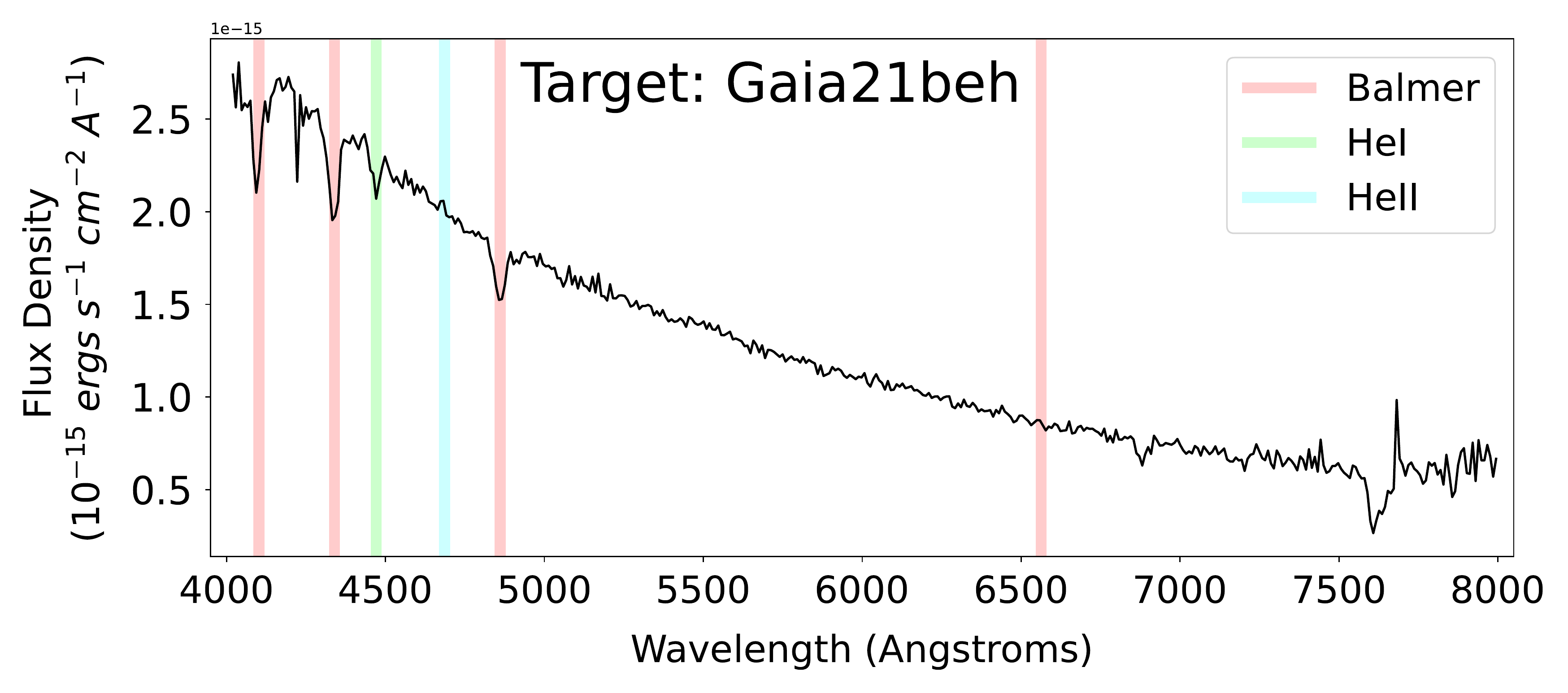}
        \label{fig:Gaia21beh}
    \end{subfigure}
\end{figure*}

\begin{figure*}\ContinuedFloat
    \centering
    \begin{subfigure}[b]{0.94\columnwidth}
        \centering
        \includegraphics[width=\columnwidth]{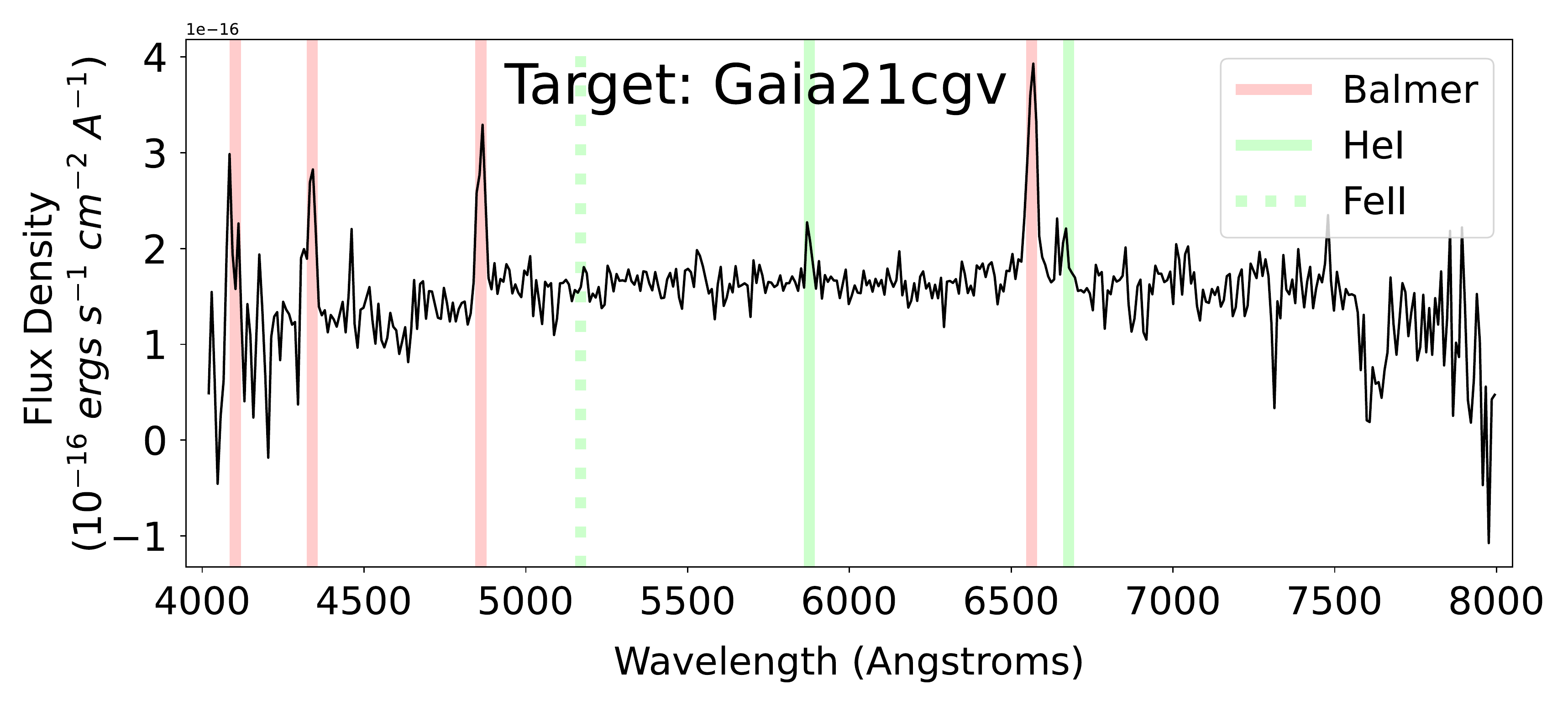}
        \label{fig:Gaia21cgv}
    \end{subfigure}
    \hfill
    \begin{subfigure}[b]{0.94\columnwidth}
        \centering
        \includegraphics[width=\columnwidth]{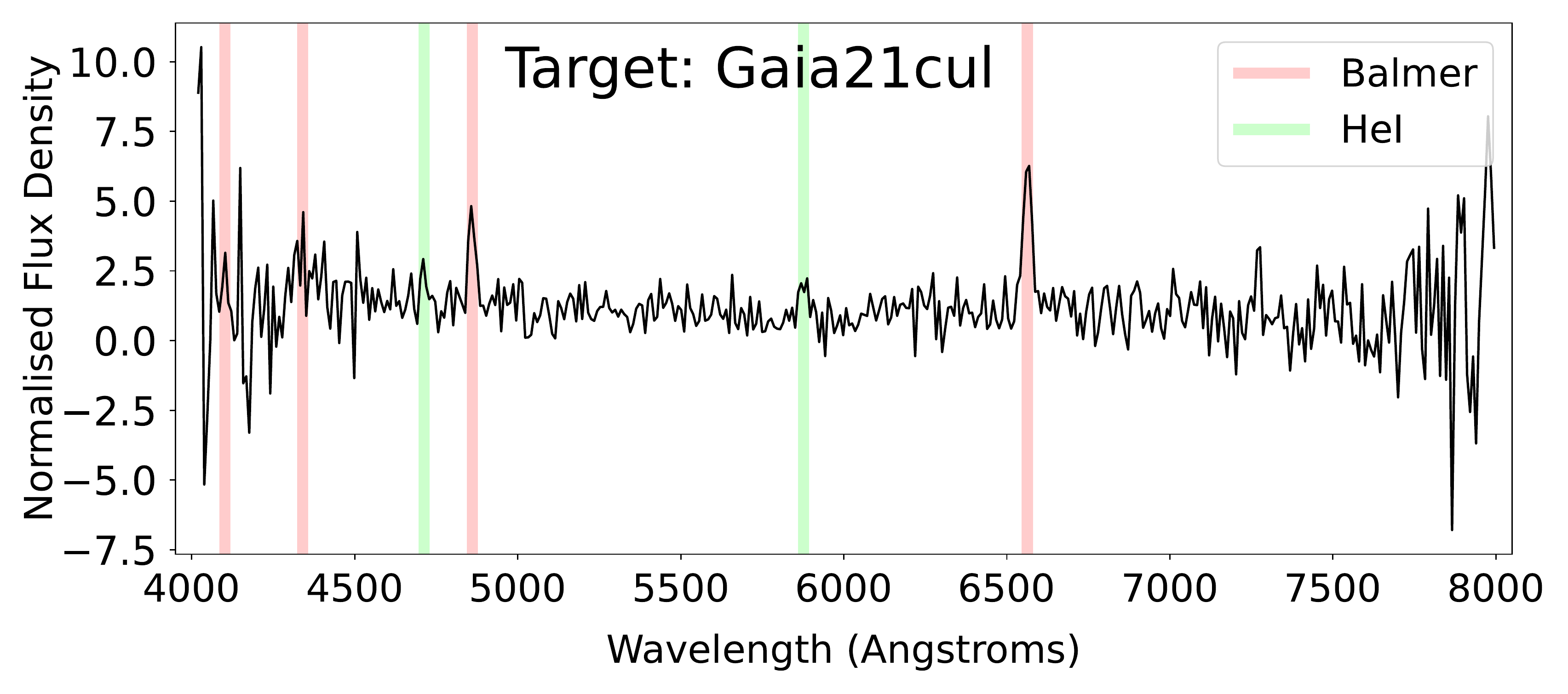}
        \label{fig:Gaia21cul}
    \end{subfigure}
    \hfill
    \begin{subfigure}[b]{0.94\columnwidth}
        \centering
        \includegraphics[width=\columnwidth]{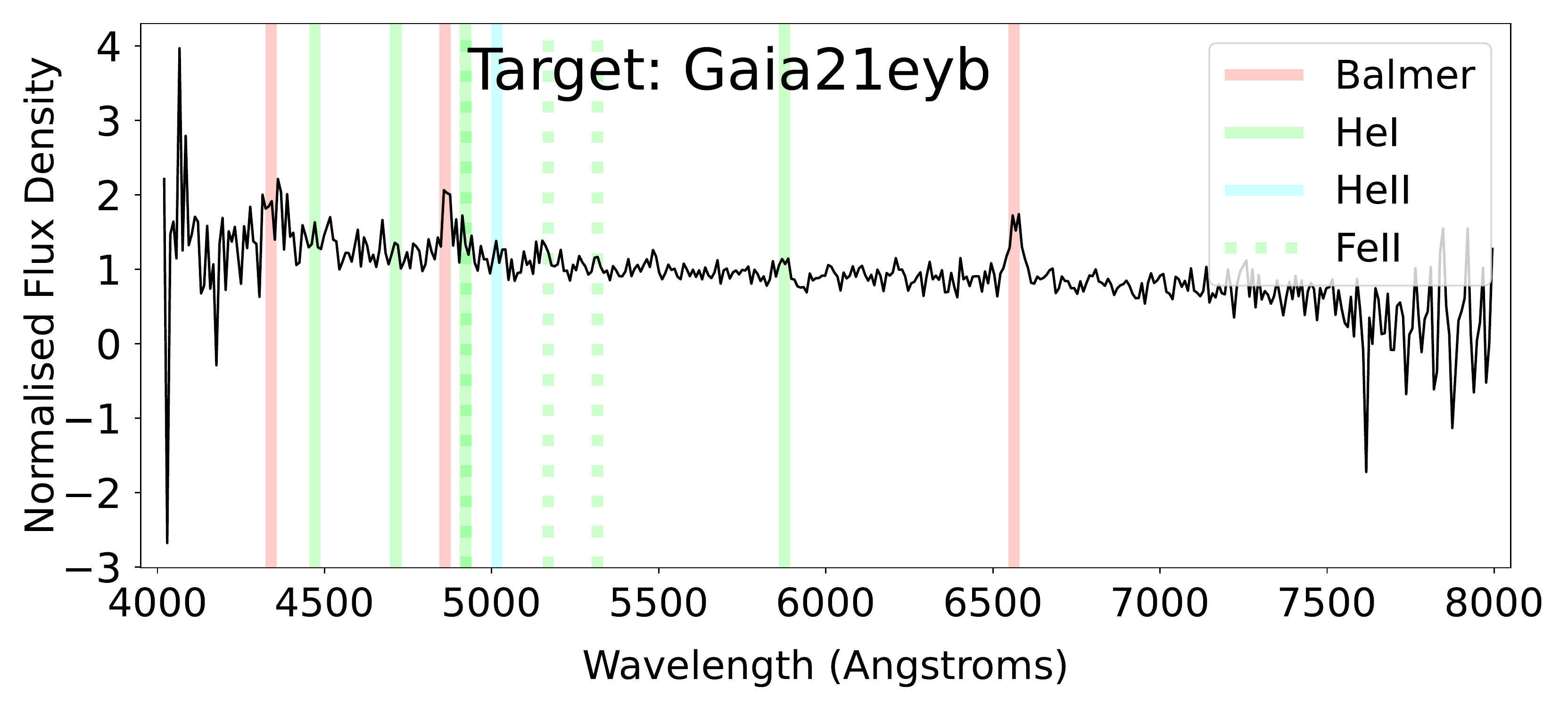}
        \label{fig:Gaia21eyb}
    \end{subfigure}
    \caption{SPRAT spectra of targets in Table \ref{tab:CV_SPRAT}. Spectral lines indicated in plots, labelled in the legend for each.}
    \label{fig:SPRAT_Spectra}
\end{figure*}

\begin{table*}
    \centering
    \begin{tabular}{llp{1.2\columnwidth}}
        \textbf{Target} & \textbf{Classification} & \textbf{Comment} \\
        \hline
        Gaia16cfn & CV (Dwarf Nova) & Clear Balmer and \ion{He}{I} emission. \ion{He}{I} $\lambda$4922 blended with \ion{Fe}{II} $\lambda$4924. Characteristic of DNe subtype.\\
        Gaia17ccv & CV (Decline from dwarf nova outburst) & CV on decline from outburst, Faint H$\alpha$ and H$\beta$ emission, \ion{He}{II} $\lambda$4686 in emission. Double peaked lines - indicative of high inclination system. \\
        Gaia17dfn & CV & Balmer and \ion{He}{I} $\lambda$4471 lines in emission\\
        Gaia18auz & CV & Clear Balmer emission with several faint \ion{He}{I} lines in emission\\
        Gaia18dgt & CV (Dwarf Nova) & Broad Balmer emission with lines of \ion{He}{I}. \ion{He}{I} $\lambda$4922 blended with \ion{Fe}{II} $\lambda$4924. Characteristic of DNe subtype. Double peaked emission, possible high inclination system\\
        Gaia18dhv & CV & Balmer, \ion{He}{I} and \ion{He}{II} in emission\\
        Gaia19bzn & CV & Clear Balmer emission; faint lines of \ion{He}{I} and \ion{He}{II} \\
        Gaia19cln & CV & Clear Balmer emission; Lines of \ion{He}{I} and \ion{He}{II} also present; \ion{He}{I} $\lambda$4922 blended with \ion{Fe}{II} $\lambda$4924 \\
        Gaia20air & CV & Clear Balmer emission; Lines of \ion{He}{I} and \ion{He}{II} also present; \ion{He}{I} $\lambda$4922 blended with \ion{Fe}{II} $\lambda$4924 \\
        Gaia20bjd & CV & Clear Balmer emission; Lines of \ion{He}{I} also present; \ion{He}{I} $\lambda$4922 blended with \ion{Fe}{II} $\lambda$4924 \\
        Gaia20cpq & CV (Dwarf Nova) & Clear Balmer emission; Lines of \ion{He}{I} and \ion{He}{II} also present; \ion{He}{I} $\lambda$4922 blended with \ion{Fe}{II}\\
        Gaia21beh & CV & Outburst spectrum. Possible very faint H$\alpha$ absorption, clear absorption in remaining Balmer lines and \ion{He}{I} $\lambda$4471, \ion{He}{II} $\lambda$4686 in emission (faint).\\
        Gaia21cgv & CV & Balmer and \ion{He}{I} emission lines, faint \ion{Fe}{II} $\lambda$5169\\
        Gaia21cul & CV & Clear Balmer and \ion{He}{I} emission lines\\
        Gaia21eyb & CV & Balmer, \ion{He}{I}, \ion{He}{II} and \ion{Fe}{II} emission lines, \ion{He}{I} $\lambda$4922 blended with \ion{Fe}{II} $\lambda$4924\\
        \hline
    \end{tabular}
    \caption{Classifications based on LT SPRAT spectroscopy of several targets labelled as 'unknown' (without a transient class assignment) within Gaia Science Alerts and predicted as CV by the RF750 model.}
    \label{tab:CV_SPRAT}
\end{table*}

\section{Conclusions and future work}
The advent of wide field synoptic surveys has revolutionised time domain astronomy with their ability to detect millions of transient event per night. The use of Machine Learning is recognised as the best method of source classification for this deluge of transient sources. Machine Learning algorithms have been applied widely to data from several surveys including CRTS and ZTF photometry. In this work we applied ML techniques to the transient stream of Gaia Science Alerts, a resource not fully explored with ML. Our focus lies in the identification of Cataclysmic Variable stars, a class of transients providing ideal laboratories for the study of accretion and binary evolution. Using features extracted from light curves of classified sources and associated metadata as input we evaluated the use of Random Forest, Adaboost, XGBoost, K Nearest Neighbours, Support Vector Machines, and a Multi Layer Perceptron in performing several tasks. These are the identification of CVs in the context of binary classification (CV or non-CV) and a 4 class task (CV, AGN, SNe, YSOs). Each of these tasks was performed with and without metadata (e.g., Gaia parallaxes and colour) during training. By comparison of the F1-score of all models across both tasks, the 4 class Random Forest model trained with both light curve and metadata based features performed the best with an F1-score of 89\% when evaluated on the test set. We applied this model to the list of unclassified targets within GSA. The model predicted 2833 of these 'unknowns' to be of the CV class. We are now undertaking a spectroscopic observing campaign to spectroscopically classify a statistically significant sample of these targets to validate our models performance. So far we have been able to spectroscopically confirm 15 target to be of the CV class.

The use of data beyond light curve features seems necessary in order to achieve classification performance close to F1=90\%. However, with light curve features alone the performance of the model compared well with other works, despite more sparsely sampled light curves. The lessons learnt during this exploration of the GSA resource and the classified targets from our spectroscopic database of targets will be useful in the next phase of our research. This will be the application of ML to the multiband high cadence light curves of the ZTF survey. 

The next phase will be an opportunity to explore methods of handling class imbalance and missing data. Class imbalance, present within our dataset (see subsection \ref{sec:Alerts}), tends to bias classifiers to recognise the oversampled class more than the undersampled class. Algorithm specific solutions exist, for example, within Random Forest one may grow each tree with the same number of targets per class by oversampling or undersampling using the bootstrap sampling process \citep{RN489}. Data augmentation methods (e.g., \citealt{RN487}) to generate new examples based on existing examples will also be explored. Our use of mean imputation for handling missing data is simple and parameter free. Whilst this method can cause biases (see subsection \ref{sec:imputation}, we deemed the exploration of several imputation methods beyond the scope of this work, though it is something we will explore in work with ZTF data. The reliability of class labels will also be important for the next research phase and once LSST becomes operational. Whilst there is confidence in the methods employed in the labelling of examples used here (see subsection \ref{sec:Alerts}), we acknowledge that labelling errors do occur. This can add noise to the dataset, deteriorating classifier performance \citep{RN490} and reducing the effectiveness of performance optimisation techniques such as hyperparameter tuning. \par
The methods employed in this work are transferable to the data available from the ZTF survey. This data should provide the necessary information to identify subclasses within the CV population and pick out rare varieties that further our understanding of binary evolution. For example, the $\sim$2 day cadence provides the sampling necessary to recognise the defining characteristics of DNe subtypes, such as the superoutbursts of SU UMa systems (e.g., \citealt{RN491}) and the standstills of Z Cam systems \citep{RN367}; and identify characteristics present in outbursting AM CVns, such as the short duration rebrightenings on the fading tail of a superoutburst \citep{RN476}. Our ability to automatically distinguish between the different CV subtypes will depend upon several factors, one of which is the quality of features. Several features used in this work have so far shown their effectiveness at class distinction, others may become more significant once computed with the higher cadence data, while the development of features geared towards the identification of specific subtypes should provide further benefit. The prevalence of a given subtype within the dataset is another factor that we expect to impact classifier performance. The sensitivity of a survey to certain CV subtypes results in the under-representation of novae, AM CVns and nova-likes compared to DNe due to the rarity of eruptions, faintness, and photometric stability, respectively. This is where the methods of handling class imbalance described in the previous paragraph will become invaluable. The methods used here and the lessons learnt will aid in our goal to separate the rare CV systems from those more common and hopefully lead to a greater understanding of binary evolution.\par

\section*{Acknowledgements}
D. Mistry acknowledges a PhD studentship from Liverpool John Moores University (LJMU) Faculty of Engineering and Technology. M.J. Darnley receives funding from UK Research and Innovation (UKRI) grant number ST/S505559/1. The Liverpool Telescope was funded by UKRI grants ST/S006176/1 and ST/T00147X/1. The Liverpool Telescope is operated on the island of La Palma by Liverpool John Moores University in the Spanish Observatorio del Roque de los Muchachos of the Instituto de Astrofisica de Canarias. We thank Dr Simone Scaringi and an anonymous reviewer for their careful reading of our original manuscript and their many insightful comments and suggestions that have helped to result in an improved final version.

\section*{Data Availability}
The full list of unclassified GSA sources that our 4 class full feature Random Forest model predicted as belonging to the CV class is provided as supplementary material online. A shortened version is shown in table \ref{tab:Predicted_CVs}.
The Liverpool Telescope spectroscopic data for the sources in Table \ref{tab:CV_SPRAT} can be acquired from \url{https://telescope.livjm.ac.uk/cgi-bin/lt_search}. Both raw and calibrated data files can be obtained by entering the object name in the appropriate field.

\begin{table}
    \centering
    \caption{List of sources list as being of 'unknown' type by GSA that our 4 class full-feature Random Forest model predicted as belonging to the CV class. The full table of 2833 sources is available as supplementary material online.}
    \begin{tabular}{ccc}
        \hline
        \textbf{Gaia Object Name} & \textbf{Right Ascension} & \textbf{Declination}\\
        \hline
        Gaia17avy & 343.52368 & 65.09345\\
        Gaia18cdn & 263.44916 & -30.53935\\
        Gaia20cjd & 223.95987 & -64.73914\\
        Gaia20eno & 271.03934 & -76.22711\\
        Gaia21bnn & 50.13837 & 6.71808\\
        Gaia20fax & 338.65765 & 58.42598\\
        Gaia20efe & 120.77764 & -17.40920\\
        Gaia19dal & 312.73430 & 31.95018\\
        Gaia21bqw & 120.07837 & -38.82770\\
        Gaia21bby & 291.18503 & -34.15307\\	
        ... & ... & ...\\
        \hline
         & 
    \end{tabular}
    \label{tab:Predicted_CVs}
\end{table}



\bibliographystyle{mnras}
\bibliography{manuscript.bib} 








\bsp	
\label{lastpage}
\end{document}